\documentclass[review,3p]{elsarticle}
\usepackage{lineno,hyperref}
\usepackage{amsmath}
\usepackage{amssymb}
\usepackage{numcompress}
\usepackage{exscale}
\usepackage[mathscr]{eucal}
\usepackage{bm,upgreek}
\usepackage{eqlist} 
\usepackage{graphicx}
\usepackage{xfrac}
\usepackage{caption} 
\usepackage{enumitem}
\usepackage{glossaries}
\biboptions{sort&compress}
\usepackage{float}
\usepackage[titletoc]{appendix}
\usepackage[section]{placeins} 
\usepackage{flafter} 
\usepackage{listings}
\usepackage{textcomp}
\usepackage{epstopdf}
\usepackage{cleveref}
\usepackage{subfig}    
\usepackage{units}

\modulolinenumbers[5]

\journal{Journal of Sound and Vibration}

\graphicspath{{Scitech_2021/Figures/}}

\begin{document}
\begin{frontmatter}

\title{A Time-Domain Linear Method for Jet Noise Prediction and Control Trend Analysis}

\author{Chitrarth Prasad\corref{abc}} \cortext[abc]{Corresponding author} \ead{prasad.141@osu.edu} \author{Datta V. Gaitonde\corref{}}

\address{Department of Mechanical and Aerospace Engineering, The Ohio State University, Columbus, OH 43210}
\ead{gaitonde.3@osu.edu}



\begin{abstract} 
Large-scale turbulent structures in the form of coherent wavepackets play a significant role in the generation of prominent shallow-angle noise radiation of jets. 
Economical prediction tools often model these wavepackets in the frequency-domain using stability modes of the mean flow.
The use of simplifying choices, such as parabolized equations and azimuthal decomposition,  provide efficient methods but can impose constraints on rate of streamwise 
variation of the mean state or geometric complexity.
The current investigation develops a time-domain linearized Navier-Stokes-based approach predicated on the mean basic state for two goals: (i) to obtain the radiated shallow-angle noise field, including that from imperfectly expanded jets containing shock trains, and (ii) to estimate  noise control trends with actuator frequency.
A previously developed implicit linearization technique repurposing native non-linear Navier-Stokes code capabilities avoids any additional constraints on nozzle geometry, while its time-domain nature facilitates control analysis through transient pulse response.
Two other integral components of the method are the sifting of linearized perturbations to isolate the acoustic component according to Doak's momentum potential theory (MPT), and subsequently Dynamic Mode Decomposition (DMD) to analyze the response in different spectral ranges.
Comparisons with well-validated Large Eddy Simulation (LES) databases in different spectral ranges show accurate model predictions for super-radiative shallow angle noise, including for hot jets from military-style nozzles.
For control frequency effectiveness, the 
transient perturbation growth characteristics are processed to delineate the distinct effects of the excitation on vorticity versus the MPT acoustic variable. 
For a specific jet with extensive published experimental data using plasma actuators, it is shown that the method correctly predicts noise amplification at lower frequencies and reduction at higher values, at the observed crossover location.
Considerations on the costs associated with the approach, which exploits linearity to extract results at multiple frequencies with each simulation, are outlined.

\end{abstract}

\end{frontmatter}





\section{Introduction \label{section:Intro}}

Jet noise forms a major component of the total noise radiated by aircraft, especially in the aft direction. 
Extreme noise levels such as those produced by low bypass-ratio supersonic military engines are responsible for noise-induced hearing loss among military personnel working in close proximity.
Jet-associated noise pollution is also a major consideration, especially in communities close to airbases and military training routes. 
The effects of aircraft noise on the health of both children and adults are well documented in literature~\cite{haines2001chronic,schmidt2013effect}. 
Public demand has motivated the exploration of noise reduction methods for existing and future tactical aircraft~\cite{martens2010practical}.  

The acoustic field is an unwanted byproduct of the mixing of exhaust gases with the ambient fluid. 
Large-scale structures in the jet plume have been established as primary contributors, especially in the case of supersonic jets~\cite{tam1971directional,sarohia1978experimental,troutt1982experiments}. 
The spatio-temporal footprint of these large-scale structures takes the form of coherent wavepackets, which constitute efficient sources of the peak downstream radiation~\cite{jordan2013wave}.
Several studies have shown that coherent wavepacket structures can be modeled in terms of linear 
stability modes of the long time-mean of the turbulent flow field. 
Different modeling approaches have been developed in the literature; the most popular among these are based on the parabolized stability equations (PSE) \cite{gudmundsson2011instability,sinha2014wavepacket}.
Although convenient, PSE are deficient in many regards -- these have been discussed in Refs.~\cite{towne2015one,rigas2017one}, which propose more accurate methods based on one-way Euler or Navier-Stokes equations.
Other techniques include global eigenvalues~\cite{schmidt2016super} and  resolvents~\cite{jeun2016input,towne2017statistical,pickering2019eddy,pickering2020resolvent}.

The application of these frequency-domain methods has generally focused on circular nozzles and perfectly-expanded operating conditions, where they are highly efficient and effective in predicting noise characteristics, especially for supersonic jets.
Application of mean-flow basic states to predict noise from imperfectly expanded jets  remain relatively scarce.
Some techniques assume slow streamwise variation of flow properties, which can be violated in the presence of shock-cells.
The exploitation of azimuthal decompositions can also introduce difficulties in conceptual extension to non-axisymmetric geometries.
Nonetheless, the success of such simple models motivates the continued vigorous development of mean-flow-based approaches for more complex situations.
In particular, they may reduce the large cost-function associated with the aeroacoustic analysis and optimization of new nozzle designs and noise control technologies, for which first-principles or cut-and-try approaches are expensive and inhibiting.

The present effort develops an approach based on linearized perturbations about the mean turbulent state to (i) predict the primary features of the shallow angle noise field applicable even to imperfectly expanded cases and (ii) estimate the effect of actuator frequency on control efforts. 
The mean turbulent state may be obtained in different ways.
One method is to use a Reynolds Averaged Navier-Stokes (RANS) solution validated by comparison with experimental data, such as in Ref.~\cite{kapusta2016numerical}.
However, since our goal is to develop the method, here we use the mean flow from well-validated LES databases; this allows different features of the perturbation field to be compared directly with the LES solution as a truth model. 
Two databases are considered: 
the first, designated Jet-A, is similar to those considered in the modeling studies of Refs.~\cite{jeun2016input,sinha2014wavepacket,pickering2019eddy}. 
Specifically, an unheated ideally-expanded Mach~1.3 jet issuing from a sleeve nozzle is considered 
as discussed in Refs.~\cite{gaitonde2011coherent,gaitonde2012analysis,unnikrishnan2016acoustic}.
The second case, Jet~B, is a more realistic over-expanded heated Mach~1.36 jet from a faceted military-style nozzle, previously examined in Refs.~\cite{prasad2019effect,prasad2020study,prasad2021steady}.
A summary of the two LES databases is provided in $\S$\ref{section:LESdatabase}.

The proposed method consists of three components.
The first adapts a recently developed
time-domain Mean Flow Perturbation (MFP) approach, originally proposed by Touber and Sandham~\cite{touber2009large} to analyze the stability of mean shock-wave turbulent boundary layer flows.
We use the variant proposed by Ranjan \emph{et al.}.~\cite{ranjan2020robust}, which provides a matrix-free method to obtain the principal stability modes of 2D and 3D mean flows on curvilinear meshes.
The MFP approach, discussed in $\S$\ref{section:MFP}, generates solutions to the linearized Navier-Stokes equations (LNSE) by modifying the native, non-linear Navier-Stokes (NS) solver used to generate the mean flow.
The linearization is implicitly performed by applying a relatively straightforward constraining body-force, computed only once and \textit{in situ}, to maintain the invariance of the base state. 
This circumvents tedious explicit linearization by repurposing existing NS solver capabilities for LNSE analyses.
As discussed in Ref.~\cite{ranjan2020robust}, the body-force term provides a measure of versatility since it facilitates decoupling of the method used to obtain the mean flow from that for the implicit LNSE; thus, dissipative shock-capturing schemes employed to generate the mean flow can be replaced by higher-order methods more suited for LNSE.
In fact, the codes employed to obtain the two LES databases are different, but the MFP procedure for both is applied using only one code.

The time-domain nature of the present approach offers crucial benefits.
First, the method is applicable to all configurations that can be computed by the underlying non-linear Navier-Stokes procedures; the solvers employed here can treat complex configurations with curvilinear meshes.
Second, the approach facilitates the analysis of transients in perturbation growth; this enables extraction of data on control effectiveness.
Since the cost of time-domain methods is typically a concern, some considerations are put forward on the expense associated with the present method.
Although the time-domain approach facilitates application to general geometries, the choice of on- and off-design axisymmetric cases is predicated on their well understood dynamics.
Separate efforts to understand and model the dynamics of non-circular jet geometries, currently being pursued with LES~\cite{chakrabarti2021representing} and PSE~\cite{rodriguez2021near}, will be addressed in a future effort.


The second component comprises sifting of the perturbation fields to separate noise-related components from primitive variables.
The LNSE perturbation evolution is initiated with random pressure fluctuations on a subset of the domain. 
Mean flow gradients subsequently induce perturbations in all other primitive quantities.
Although pressure and other primitive variables are convenient for analysis since they are directly obtained from  experiments or simulations, they are often not optimal for extraction of acoustic-related information.
For example, pressure includes both hydrodynamic and acoustic components; the former display vortical and convective characteristics, while the latter are irrotational and radiating.
Techniques to separate the convecting from radiating pressure fluctuations based on phase speeds of pressure fluctuations, using methods such as reduced-order models or spectral techniques, may be found Refs.~\cite{tinney2008near,du2015separation,mancinelli2018hydrodynamic}.

In the present work, we adopt Doak's Momentum Potential Theory (MPT)~\cite{doak1989momentum}, presented in $\S$\ref{section:Doak}, to extract a pure acoustic component.
This robust, physics-based method exactly decomposes  momentum density ($\rho \textbf{u}$) fluctuations, regardless of magnitude, into its constituent hydrodynamic, acoustic and thermal components, also designated Fluid Thermodynamic (FT) modes. 
Indeed, compared to pressure fluctuations, the acoustic FT component exhibits wave-packets with superior spatio-temporal coherence and radiative efficiency because Doak's MPT effectively filters out the turbulent and convecting high-energy fluctuations by imposing the physical constraints of irrotationality and isentropy~\cite{unnikrishnan2019acoustically,prasad2019modal}. 
In the far-field, where acoustic sources are negligible, the acoustic component recovers the desired scaled pressure fluctuations.
The advantages of such sifting has been previously discussed to analyze LES fluctuations~\cite{unnikrishnan2016acoustic,prasad2020study}
 and to derive constants in frequency-based wavepacket models~\cite{unnikrishnan2019acoustically}.
 Thus, the use of the acoustic FT component for comparison of MFP with LES provides a more effective examination of the noise radiating mechanisms.
 This component is also crucial in establishing control characteristics with actuator frequency as discussed further below.

The final component of the approach, employed to post-process the sifted time-domain LNSE perturbations, is the traditional Dynamic Mode Decomposition (DMD)~\cite{schmid2010dynamic}.
The method extracts frequency-specific modal information.
When applied to the MPT acoustic component of MFP-obtained perturbations and LES fluctuations, it provides a clear strategy for comparison of aspects related specifically to noise prediction analyses.
On the other hand, when applied to the transient  impulse response, DMD yields data on growth rates of different quantities, which can then be interpreted for control purposes as shown in this work.
The noise prediction results are presented in $\S$\ref{sec:noiseprediction}.
The impulse random pressure fluctuation response is obtained with LNSE using the MFP approach.
The resulting perturbations, which are not restricted to pressure alone, are then sifted with MPT to extract the acoustic component.
The robustness and insensitivity of the solution to different parameters inherent to the method, including forcing region, initial perturbation amplitude, and mesh resolution are demonstrated.
For verification, comparisons between the LNSE and the original LES are presented at different frequencies.
Finally, the noise spectra from the LNSE are 
 compared with those from the LES to demonstrate the proper variations away from the jet shear layer, focusing as mentioned earlier, on the shallow angle noise.

The second thrust, analyzing control trends with frequency of excitation, is based on predicting the relative growth of acoustic versus vortical responses to perturbation frequency in the peak radiation direction.
Since the noise field is directly related to the spatio-temporal signatures of large-scale structures in the jet plume, noise reduction techniques seek to enhance fine-scale mixing between the jet plume and the ambient air to break up acoustically efficient large-scale structures in the flow.
The present approach, $\S$\ref{section:Control} provides a relatively direct method to extract information about the nature of the response to excitation.
For this, the sifted and primitive variables are subjected to DMD to extract the Ritz values~\cite{schmid2010dynamic} of linearized perturbation growth.
Specifically, the relative growth rates of vorticity 
and the MPT acoustic component are compared.
The conditions chosen are again motivated by the need to compare with experimental observations.
For this, the extensive data available on the response of Jet-A to plasma-based actuation~\cite{samimy2010acoustic} are leveraged.
Concluding remarks are put forth in $\S$\ref{conclusion}.

\section{LES databases \label{section:LESdatabase}}
\subsection{Jet-A: Unheated perfectly-expanded Mach 1.3 jet}
The first database is of an unheated perfectly-expanded Mach~1.3 jet, referred to as Jet-A.
The database has been extensively  validated with experimental data and the flow and acoustic dynamics have been discussed in several prior publications.
Examples include extraction of noise generation components~\cite{unnikrishnan2016acoustic} and the effect of plasma actuators for noise reduction~\cite{gaitonde2011coherent,gaitonde2012analysis}.

The database was generated by solving the full 3D compressible Navier-Stokes equations in generalized curvilinear coordinates.
The jet exit conditions are $U_j = 391 \unit{m/s}, T_j = 224 \unit{K} \text{ and } \rho_j=1.567 \unit{kg/m^3}$.
Flow quantities are non-dimensionalized using these as reference values for  velocity, temperature and density respectively. 
The nozzle exit diameter, $D=0.0254$m is used as the reference length and the pressure is normalized by $\rho_j {U_j}^2$. 
The ambient temperature is $T_\infty = 300 \unit{K}$. 
The inflow is specified through a short constant area sleeve in the manner specified by Bogey and Bailly~\cite{bogey2010influence}.
The nozzle exit is located at $x/D \sim 0.17$.

A third-order upwind biased Roe scheme~\cite{roe1981approximate} in conjunction with the van Leer harmonic limiter~\cite{van1979towards} is employed for the spatial discretization of the inviscid fluxes. 
The viscous fluxes are computed using a second-order central differencing scheme, and time integration is achieved with an implicit second-order diagonalized~\cite{pulliam1981diagonal} Beam Warming scheme~\cite{beam1978implicit}. 
A detailed description of mesh resolution studies and boundary conditions can be found in Ref.~\cite{gaitonde2011coherent}. The present database uses a cylindrical structured mesh containing $523$, $526$ and $125$ points in the axial, radial and azimuthal directions respectively. 
The mesh is clustered in the radial and axial directions near the centerline and the nozzle walls with gradual stretching towards outer boundaries. 

\begin{figure}[!h]
    \centering
    \subfloat[]{\includegraphics[width=0.7\linewidth]{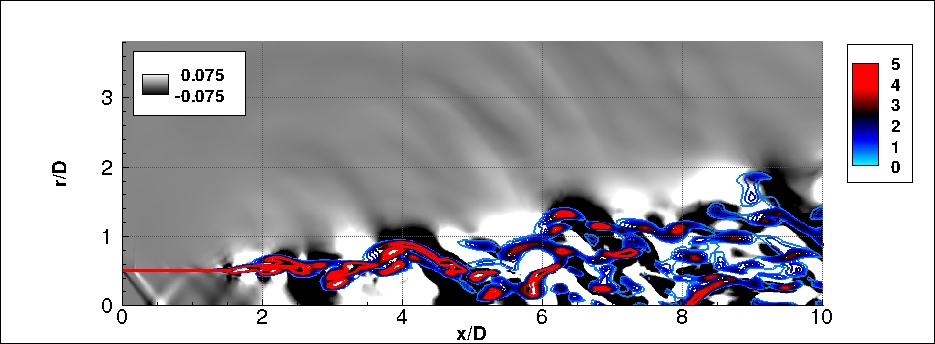}} \\
    \subfloat[\label{fig:GE404dengrad}]{\includegraphics[width=0.7\linewidth]{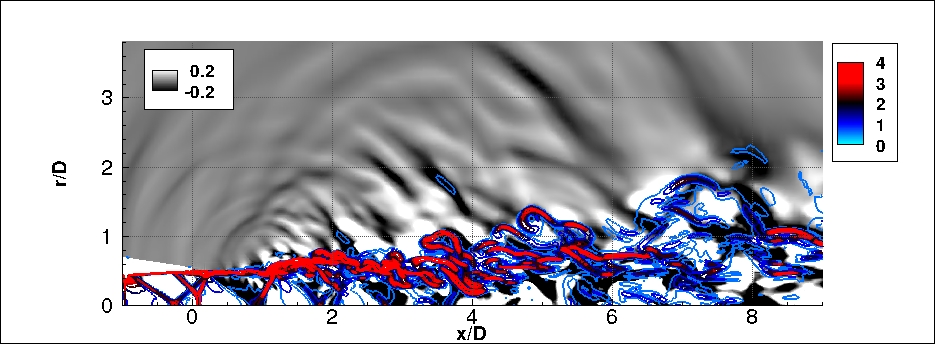}}
    \caption{Instantaneous LES flow-field in a 2D streamwise plane. (a) Jet-A: Vorticity contours superimposed on velocity dilatation contours and (b) Jet-B: Density gradient contours superimposed over velocity dilatation contours.  }
    \label{fig:M1_3LES}
\end{figure}

\begin{figure}[!h]
    \centering
\subfloat[]{\includegraphics[width=0.5\linewidth]{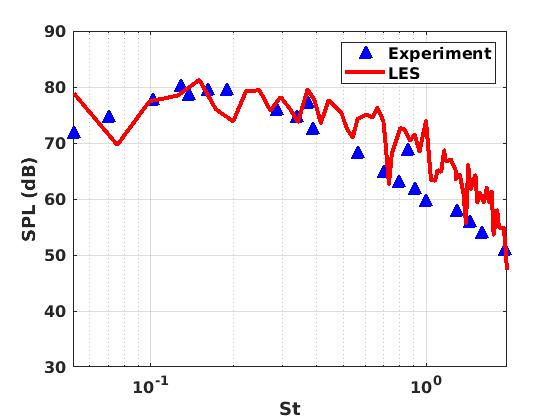}}
\subfloat[\label{fig:GE404spectra}]{\includegraphics[width=0.5\textwidth]{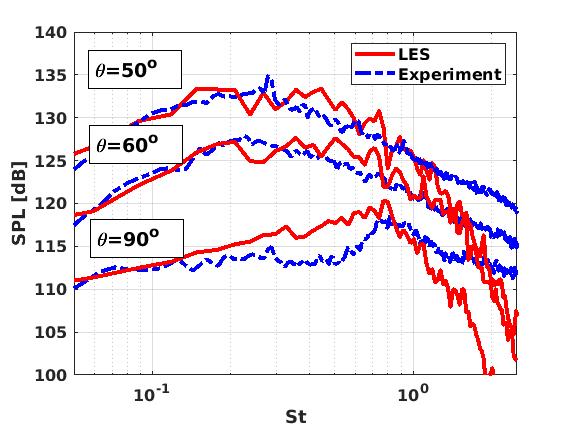}}
    \caption{
    Far-field sound pressure level for (a) Jet-A at $94$ jet diameters from the jet exit at $\theta=30^\circ$ from the jet downstream axis, (b) Jet-B at $100$ jet diameters at three different polar angles. The peak noise radiation for Jet-A occurs at $\theta_{peak}=30^\circ$ and for Jet-B occurs at $\theta_{peak}=47.5^\circ$.}
    \label{fig:compareexp}
\end{figure}

Figure~\ref{fig:M1_3LES}(a) shows vorticity contours (color) superimposed on velocity dilatation contours (gray-scale) at a representative time-step on an azimuthal plane. 
The vorticity contours track the growth and breakdown of the jet shear layer as it mixes with the ambient fluid, whereas the background dilatation contours provide a qualitative picture of the generated acoustic waves. 
As expected for a supersonic jet, the acoustic waves display a strong downstream radiation pattern. 
This is representative of the superdirective nature of the jet. 
Although the database has been extensively validated previously \cite{gaitonde2011coherent,gaitonde2012analysis,unnikrishnan2016acoustic}, a comparison of far-field pressure spectrum in the peak noise radiation direction ($\theta=30^\circ$ from the jet downstream axis) with experimental measurements is provided in Fig.~\ref{fig:compareexp}(a) for completeness. 
The LES predictions are obtained using the Ffowcs-Williams Hawkings (FW-H) analogy~\cite{ffowcs1969sound}, with the far-field observer located at $94$ jet diameters from the jet exit. 
Overall, the LES predictions are in reasonable agreement with the experimental measurements. 
Other results, including comparisons of  the mean centerline velocity obtained from long time averaging of the LES with the PIV and velocity fluctuation measurements of Ref.~\cite{samimy2007active} may be found in Ref.~\cite{gaitonde2011coherent}. 

\subsection{Jet-B: Heated Over-expanded Mach~1.36 jet}
The second LES database used in this study represents a model-scale GE-F404 nozzle operating at a Nozzle Pressure Ratio (NPR) of $3.0$ and a Total Temperature Ratio (TTR) of $2.5$. 
This corresponds to an over-expanded operating condition at $M_j=1.36$. 
The database has also been validated and used extensively for the development and understanding of fluid injection-based noise reduction devices in Refs.~\cite{prasad2019effect,prasad2020study,prasad2021steady}.  
The database was generated with the STAR-CCM+ code using a dynamic Smagorinsky subgrid scale model on trimmed hexahedral meshes.
The use of a commercial code combined with the off-design nature of the jet makes this an important test case to assess the robustness and feasibility of the proposed method for a practical jet noise problem relevant to industry.
The LES data is non-dimensionalized using ambient values of the speed of sound $(c_\infty=340.17 \unit{m/s})$, temperature $(T_\infty = 288 \unit{K})$ and density ($\rho_\infty=1.226 \unit{kg/m^3}$). 
The nozzle exit diameter, $D=0.0254 \unit{m}$ is used as the reference length, whereas the pressure is normalized by $\rho_\infty {c_\infty}^2$.  
Unlike Jet-A, the physical nozzle is a part of the LES domain and the nozzle exit is located at $x/D =0$.
The nozzle design consists of a sharp throat, with a faceted diverging section comprising of $12$ seals and $12$ flaps to facilitate for area adjustment in operational nozzles.
A detailed description of the nozzle geometry is provided in Ref.~\cite{kuo2012acoustic}.
In the LES domain, the nozzle exit area ratio is fixed at $1.295$ to facilitate a comparison with available experimental data~\cite{morgan2017further}.
The flow through the nozzle is simulated by specifying the values of stagnation pressure and temperature at the nozzle inlet. 

A hybrid 3rd-order MUSCL/central differencing scheme~\cite{west2015jet} in conjunction with the min-mod limiter is used for spatial discretization of convective fluxes, whereas a second-order implicit scheme with sub-iterations is employed for time integration. 
The mesh consists of approximately $18.1$ million volumes with prismatic cells near the solid boundaries where $y^+< 1$. 
Other details of the discretization procedure,  mesh distribution and  grid resolution studies may be found in Refs.~\cite{prasad2019unsteady,prasad2020study}. 

Figure~\ref{fig:M1_3LES}(b) shows the instantaneous density gradient contours superimposed over velocity dilatation contours for this jet in a 2D streamwise plane. 
The shock-cell structures due to the over-expanded nature of the jet are clearly visible from the density gradient distribution. 
The dilatation contours, on the other hand, exhibit a strong downstream Mach wave radiation signature characteristic
of heated jets. 
As expected, the downstream radiation is much more intense when compared with Jet-A (Fig.~\ref{fig:M1_3LES}(a)) because of the supersonic convection of large-scale structures in the jet plume with respect to ambient.  
Additionally, the sideline and upstream radiation due to the interaction of the jet shear layer with the shock-cells is also captured by the dilatation contours.

The near-field LES results are extrapolated to the far-field at $100$ jet diameters from the nozzle exit using the FW-H analogy, closely following the experimental procedure to obtain the far-field microphone signals as described in Ref.~\cite{prasad2020coupled}. 
Figure~\ref{fig:compareexp}(b) compares the far-field predictions of the LES with the experimental measurements at three different polar angles. 
At downstream locations, where Mach waves are dominant and are the focus of the present study, the LES predictions accurately capture the shape of the spectra up to the grid cut-off \emph{St} values ($\approx 1$)~\cite{prasad2021steady}.
At sideline locations, the LES accurately resolves the peak associated with the broadband shock-associated noise (BBSAN).


\section{Numerical components of the approach}
In this section, we summarize the Mean Flow Perturbation (MFP) and data sifting with Momentum Potential Theory (MPT), emphasizing aspects of interest to the present work.
The third component, Dynamic Mode Decomposition (DMD), is used in its standard form as in Ref.~\cite{schmid2010dynamic} and is thus omitted for brevity.
\subsection{Mean Flow Perturbation (MFP) Formulation\label{section:MFP}}
As noted in the introduction, the perturbations are propagated with an implicit approach which repurposes the baseline NS solver to yield LNSE solutions about the mean flow basic state.
When used to explicitly extract the stability modes, the method represents a matrix free approach~\cite{ranjan2020robust}; in the present work however, the interest is in the time-accurate perturbation evolution for noise-prediction.
For completeness, the implicit linearization may be summarized as follows.
The standard NS equations may be written as:
\begin{equation} \label{eqn:MFP}
    \frac{\partial \textbf{Q}}{\partial t} = {\textbf{F}\left( \textbf{Q}\right)} + \textbf{B}_\text{f},
\end{equation}
where $\textbf{B}_\text{f}$ is a suitable body-force as defined below and
$\textbf{Q}$ is the state vector.
Rewriting Eq.~(\ref{eqn:MFP}) as
\begin{equation}\label{eqn:MFP2}
    \frac{\partial \left({\overline{\textbf{Q}} + \textbf{Q}'}\right)}{\partial t} = {\textbf{F}\left( \overline{\textbf{Q}} + \textbf{Q}'\right)} + \textbf{B}_\text{f},
\end{equation}
and using Taylor series about the base flow, $\overline{\textbf{Q}}$, Eq.~(\ref{eqn:MFP}) can be further rearranged as:
\begin{equation} \label{eqn:MFP_expanded}
    \frac{\partial \textbf{Q}'}{\partial t} + \left( \frac{\partial \overline{\textbf{Q}}}{\partial t} - \textbf{F}\left( \overline{\textbf{Q}}\right) - \textbf{B}_\text{f}\right) = \frac{\partial \textbf{F}}{\partial \textbf{Q}}\textbf{Q}' + \frac{\partial ^2 \textbf{F}}{\partial \textbf{Q}^2}\textbf{Q}'^2 + ... \text{(higher order terms)}.
\end{equation}
Thus, if the  standard NS equations are solved with a body-force given by
\begin{equation} \label{eqn:MFP_Bf}
    \textbf{B}_\text{f} = \frac{\partial \overline{\textbf{Q}}}{\partial t} - \textbf{F}\left( \overline{\textbf{Q}}\right)
\end{equation}
and the perturbations  $(\textbf{Q}')$ are small enough so that the contribution from the non-linear terms is negligible, the procedure successfully provides the evolution of perturbation according to the LNSE on the base flow $(\overline{\textbf{Q}})$.
The approach effectively linearizes about the conserved variables (cf. Ref.~\cite{karban_ambiguity_2020}).

The basic state for MFP is obtained for each LES by time-averaging over a period of $300 D_j/U_j$ seconds.
The evolution of the perturbation flow-field is obtained after specification of an initial distribution.
Different initial condition possibilities have been discussed in Ref.~\cite{ranjan2020robust}, depending on the objective.
The choices for the present cases, and the underlying rationale and sensitivity, are detailed in the context of each problem for noise prediction ($\S$\ref{sec:noiseprediction}) as well as the control analysis ($\S$\ref{section:Control}).
The linearized solution in the form of snapshots suitable for post-processing is then obtained by marching forward in time, using the NS equations with the modified body-force term in Eq.~(\ref{eqn:MFP_Bf}).

As noted earlier, even though the LES databases are generated using two separate NS solvers, the same structured code employed for Jet-A LES is also used for Jet-B to perform the MFP.
For this,  the mean flow-field for Jet-B is interpolated onto a nominal structured cylindrical mesh containing $651 \times 251 \times 125$ mesh points.
Regardless of the numerical procedure used to obtain the basic states, the perturbation field is evolved for both jets with a third-order Roe scheme for spatial discretization and a third-order Runge-Kutta scheme for time integration; 
these are selected due to their widespread availability 
in most commercial and in-house solvers. 
The implicit nature of the MFP procedure offers significant flexibility in the specification of boundary conditions as they can be applied to either the perturbation field or the total flow quantities~\cite{ranjan2020robust}. 
In the present work, standard no-slip boundary conditions are applied to the perturbations at the nozzle walls, whereas at outer boundaries, characteristic boundary conditions are enforced on the total flow variables.

The costs associated with the time-domain approach relative to those using frequency-based methods depend on the objectives. 
Some considerations on the computational expense of the current approach are now presented.  
The effort to obtain the body-force $\textbf{B}_\text{f}$ of Eq.~(\ref{eqn:MFP_Bf}) is negligible, since it is only computed once. 
Thus, the cost of each iteration per mesh point for MFP is similar to that for the LES.
The cost of sifting is also similar to that required for the LES.
However, several factors substantially lower the cost of MFP versus LES by orders of magnitude.
(i) The mesh resolution required to support the mean flow is far smaller than that for the LES, since the finer convecting scales are averaged out.  
A practical demonstration of this is presented in $\S$\ref{sec:noiseprediction}.
(ii) The linear nature of MFP evolution facilitates the use of high-order spatio-temporal methods which further reduces costs.
(iii) The MFP equations are only marched for a small fraction of the time required to obtain statistically converged LES data. 
(iv) A single time-domain solution provides snapshot data for the entire range of frequencies of interest (assuming the time-step-size is chosen appropriately to resolve these temporal scales).
If superdirectivity is of interest, with focus on the axisymmetric $m=0$ mode, the simulation becomes essentially two-dimensional.
As an example, the Jet-A LES calculation required approximately $90{,}000$ CPU hours with 768 Intel\textsuperscript{\textregistered} Xeon\textsuperscript{\textregistered} 8268s Skylake (2.9GHz) processors ($\sim 120$ hours of wall-clock time).
In contrast, the corresponding MFP-based results below were obtained in $24$ CPU hours ($\sim 30$ minutes wall-clock) on a Linux desktop workstation with 48 Intel\textsuperscript{\textregistered} Xeon\textsuperscript{\textregistered} E7-4809 v4 (2.10 GHz) processors. 



\subsection{Data sifting for noise radiation studies\label{section:Doak}}
The implicit LNSE provides the evolution of fluctuations of primitive variables.
Although $p'$ is the quantity of ultimate interest in the far-field, its value in the near-field is not purely representative of propagating acoustic content.  
Rather, it also includes a hydrodynamic component~\cite{bogey2007experimental}, whose influence adversely affects noise prediction techniques~\cite{unnikrishnan2019acoustically}.
The extraction of a variable representing acoustic radiating components from the primitive fluctuations is highly advantageous for developing aeroacoustic theories and inform noise control techniques. 

A robust physics-based method for simulation data that contain information on fluctuations of all variables is Doak's MPT, which separates radiating and the non-radiating spatio-temporal components.
A summary is provided for completeness; the theory may be found in Ref.~\cite{doak1989momentum}, with application to jet noise LES analysis in Refs.~\cite{unnikrishnan2016acoustic,prasad2020study}.
The key step is a decomposition of the momentum density $(\rho \textbf{u})$  into FT modes representing hydrodynamic, acoustic and thermal components:
\begin{equation} \label{Doak1}
\rho \textbf u = \bar{\textbf B} + \textbf B' - \nabla \psi'
\end{equation}
where
\begin{equation} \label{Doak2}
\nabla \cdot\bar{\textbf B}=0, \hspace{0.2in }\nabla \cdot\textbf B'=0
\end{equation}
Here $\bar{\textbf B}$ is the mean solenoidal component, $\textbf  B'$ is the fluctuating solenoidal component and $\psi'$ is the fluctuating irrotational scalar potential. 
Eqs.~(\ref{Doak1}) and (\ref{Doak2}) when used with the mass conservation equation yield,
\begin{equation}\label{Doak_irr}
\nabla^2 \psi'=\frac{\partial \rho'}{\partial t}
\end{equation}
The irrotational scalar potential can be further split into acoustic and thermal FT modes: $\psi'=\psi'_a + \psi'_t$, which are described by,  
\begin{equation} \label{Doak_irr2}
\nabla^2 \psi_a'=\frac{1}{c^2}\frac{\partial p'}{\partial t}, \hspace*{0.2 in} \nabla^2 \psi_t'=\frac{\partial \rho'}{\partial S}\frac{\partial S'}{\partial t}
\end{equation}
where $\psi'_a$ and $\psi'_t$ are the acoustic and thermal components of the irrotational momentum density, the entropy $S$ is calculated relative to the jet exit conditions and is given by,
\begin{equation}
    S=C_p [\log(T/T_{j}) - R\log(p/p_{j})]
\end{equation}
Here $C_p$ is the specific heat at constant pressure and $R$ is the ideal gas constant. 

The acoustic FT mode is extracted from the LES and MFP data by solving the Poisson equation for $\psi'_a$ at every time instant. 
The Poisson equation is discretized to second-order accuracy in generalized curvilinear coordinates and solved using the BiCGSTAB algorithm~\cite{van1992bi}.
The boundary conditions are similar to the ones used previously to to analyze LES fluctuations for both Jet-A~\cite{unnikrishnan2016acoustic} and Jet-B~\cite{prasad2020study}.
The outer boundaries are assumed to be purely acoustic; this results in $(\rho \textbf{u})'=-\nabla \psi'_a$, which is integrated along the boundaries to provide Dirichlet boundary conditions to the Poisson solver. 
A zero-gradient Neumann condition for $\psi'_a$ is enforced at the centerline due to the axisymmetric nature of the jets considered.

\section{Noise Prediction} \label{sec:noiseprediction}

\subsection{Growth of Perturbations}

\begin{figure}[!h]
    \centering
    \subfloat[]{\includegraphics[width=0.65\textwidth]{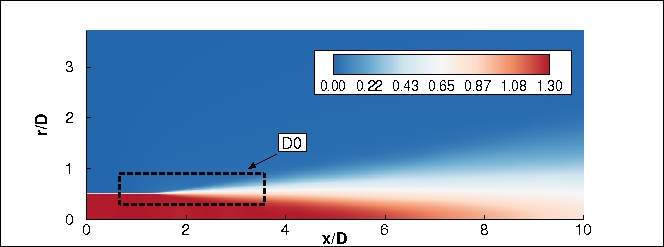}} \\
    \subfloat[]{\includegraphics[width=0.65\textwidth]{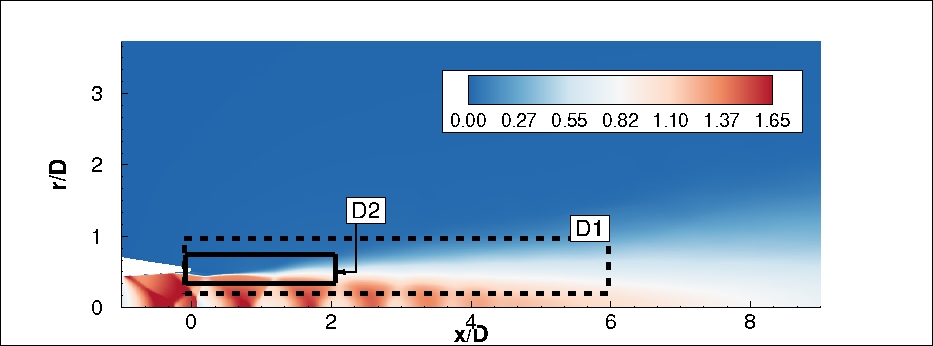}}
    \caption{Mean Mach number contours showing the regions of initial perturbation for (a) Jet-A and (b) Jet-B.}
    \label{fig:InitialPertub}
\end{figure}

The Mach number field of the time-mean flow for Jet-A and Jet-B are shown in Fig.~\ref{fig:InitialPertub}.
For the latter, a shock train is initiated inside the nozzle because of its military-style design.
In the present case, since Jet-A and Jet-B are both axisymmetric, and the focus is on superdirectivity analysis, a 2D 
slice suffices because of the dominance of the axisymmetric mode~\cite{jeun2016input}. 

The first step is the specification of the initial perturbation field. 
We take guidance from efforts to obtain stability modes such as those for supersonic jets in Ref.~\cite{rodriguez2013inlet} and flows past airfoils and cylinders in Ref.~\cite{ranjan2020robust}.
The wavepacket dynamics are governed by the Kelvin-Helmholtz (K-H) instability in the jet shear layer~\cite{rodriguez2013inlet}. 
In the context of input-output systems, the K-H instability is associated with a white-noise forcing in the thin shear layer in the vicinity of the nozzle exit~\cite{jeun2016input,schmidt2018spectral}.
In the present investigation therefore, for both jets, we consider random pressure forcing of the form
\begin{equation} \label{eqn:forcing}
    p'(\textbf{x},t)=\epsilon_0 \text{RAND}(\textbf{x})\delta(t-t_0) p_\infty.
\end{equation}
Here $\text{RAND}(\textbf{x})$ generates a random normalized distribution in space with values between $-1$ and $1$. The perturbations are scaled by the freestream value of pressure ($p_\infty$) and an initial perturbation magnitude $\epsilon_0$. 
The use of the Dirac-delta function ensures that the forcing is only applied at time $t_0$ \textit{i.e.,} the perturbation is an initial condition and yields an impulse response. 
This choice 
aids in establishing the linear independence of snapshots which can be crucial for stability mode extraction~\cite{ranjan2020robust} and growth rate determination ($\S$\ref{section:Control}).
The 
uncorrelated nature of the initial forcing ensures that the range of frequencies encompasses 
frequencies of known principal
modes and is consistent with K-H forcing in previous studies of jet noise with input-output systems.
In order to assess the sensitivity of the MFP solution to the perturbation magnitude $\epsilon_0$ and the forcing location, four test cases are presented in Table~\ref{tab:MFPCases}.
The first three cases, Pert01 through Pert03, are computed with an initial perturbation of $\epsilon_0=10^{-5}$ for both jets.
The sub-domains where this initial perturbation is applied are marked as D0 for Jet-A, while for Jet-B, two regions D1 and D2 are considered as shown in Fig.~\ref{fig:InitialPertub}.
The effect of perturbation magnitude and mesh resolution is examined using a fourth test-case, Pert04 which uses an initial perturbation with $\epsilon_0=10^{-6}$ in the sub-domain D2 and is marched forward on a coarser mesh obtained by extracting every alternate point in the streamwise and radial directions of the LES mesh. 
Further ramifications of these choices of forcing location, mesh resolution and the $\epsilon_0$ value are discussed in context below.

\begin{table}[]
    \centering
     \caption{List of test cases for Jet-A and Jet-B with MFP.}
    \begin{tabular}{|c|c|c|c|c|}
    \hline
         Case No. & Basic State &  Sub-Domain & $\epsilon_0$ & Mesh\\ \hline
         Pert01 & Jet-A & D0 & $10^{-5}$ & 523x526 (Same as LES) \\
         Pert02 & Jet-B & D1 & $10^{-5}$ & 651x251 (Same as LES) \\
         Pert03 & Jet-B & D2 & $10^{-5}$ & 651x251 (Same as LES) \\
         Pert04 & Jet-B & D2 & $10^{-6}$ & 326x126 ($1/4^{\text{th}}$ of LES) \\\hline
    \end{tabular}
   
    \label{tab:MFPCases}
\end{table}



\begin{figure}[!h]
    \centering
    \subfloat[]{\includegraphics[width=0.5\linewidth]{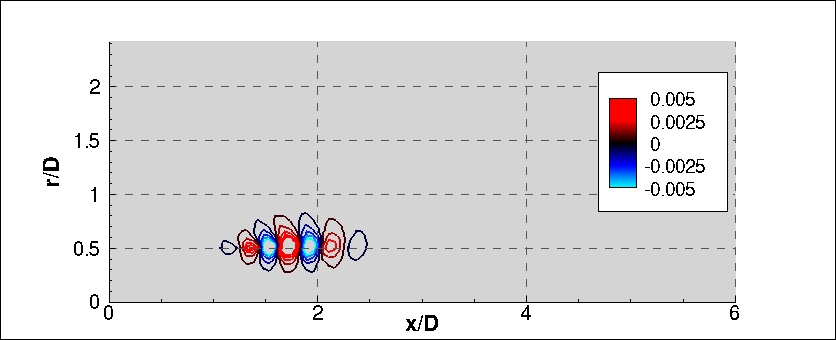}}
     \subfloat[]{\includegraphics[width=0.5\linewidth]{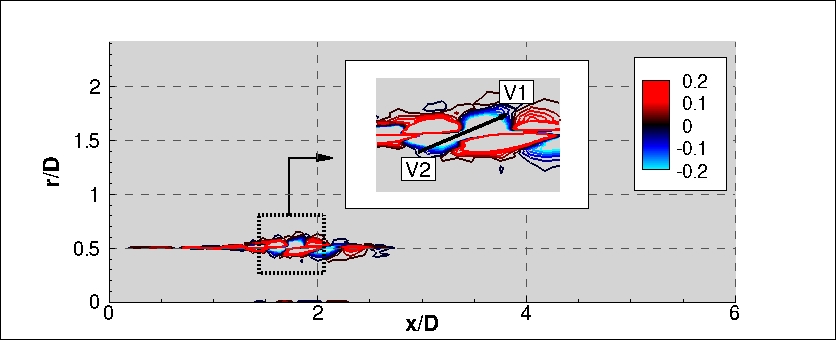}} \\
    \subfloat[]{\includegraphics[width=0.5\linewidth]{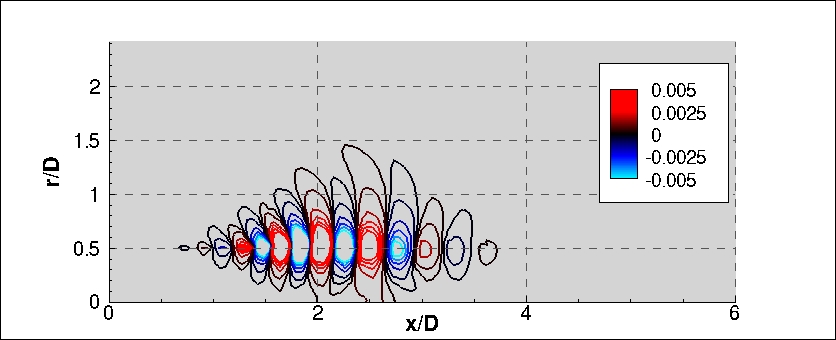}} 
    \subfloat[]{\includegraphics[width=0.5\linewidth]{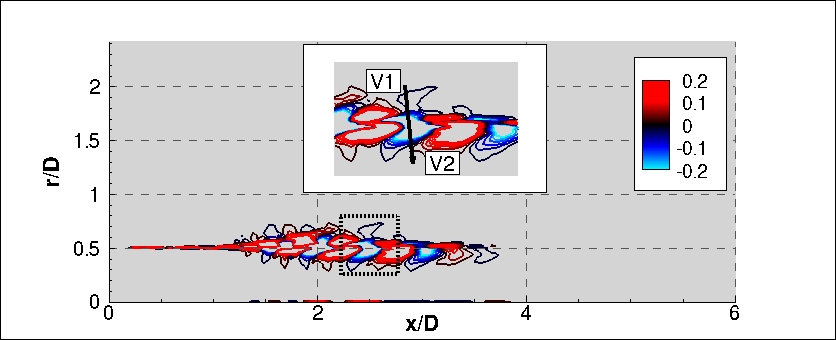}}  \\
    \subfloat[]{\includegraphics[width=0.5\linewidth]{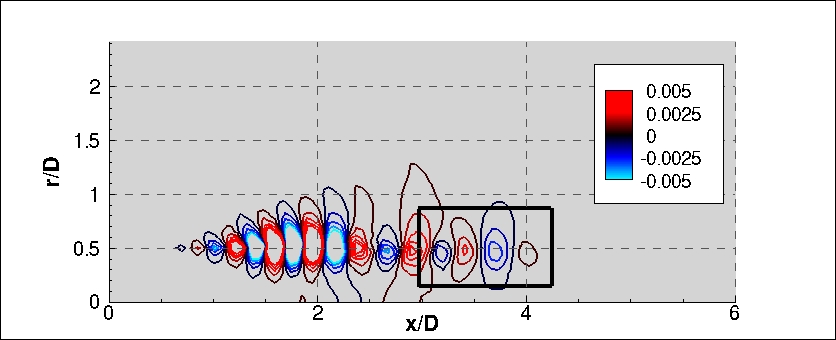}}
     \subfloat[]{\includegraphics[width=0.5\linewidth]{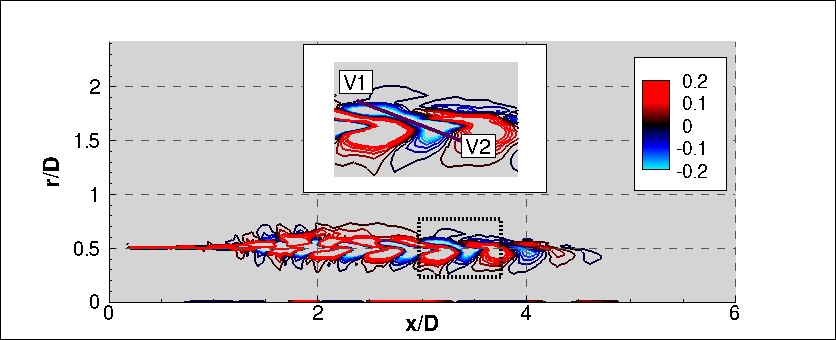}} 
    \caption{Jet-A: Growth of pressure (left) and vorticity (right) perturbations with time: $t=3.525$ (top), $t=4.775$ (middle), and $t=6.025$ (bottom). Here $t$ represents the non-dimensional time.}
    \label{fig:PressMFP}
\end{figure}

All test cases 
are marched forward in time with a non-dimensional time-step $\Delta t=5 \times 10^{-4}$.
This $\Delta t$ value is found to be sufficient to avoid convergence issues while keeping a short run-time.
A total of $1{,}200$ snapshots are sampled at an effective $\Delta t=0.025$. 
This corresponds to a minimum frequency resolution of \emph{St}$=0.033$ and a Nyquist frequency cut-off of \emph{St}$=20$. 
At this sampling rate, a typical frequency of interest, \emph{St}$=0.3$, is resolved with 133 snapshots  (nine cycles).
Figures~\ref{fig:PressMFP}(a), (c) and (e) show the evolution of pressure perturbations with time at three representative time instances for Jet-A. 
At any instant, the pressure perturbation field has a wavepacket structure 
resembling
the grayscale dilatation contours of the full LES solution displayed in Fig.~\ref{fig:M1_3LES}(a). 
The pressure field in Fig.~\ref{fig:PressMFP} contains content that represent the radiating component of the flow,
as well as
hydrodynamic content, similar to that observed in LES fluctuations~\cite{unnikrishnan2019acoustically}.
Moreover, as mentioned earlier, the effect of gradients in the mean flow is to generate perturbations in all other quantities.
For example, Figs.~\ref{fig:PressMFP}(b), (d) and (f) show the evolution of vorticity perturbations at 
corresponding
time instants.
Since the design aim of many noise reduction technologies is to enhance mixing to break up the acoustically efficient large-scale structures in the jet plume, the growth of vorticity perturbations is helpful as one possible measure of the mixing introduced by the initial perturbations.
This connection is leveraged for analysis of control as further elaborated in $\S$\ref{section:Control}.

An examination of the evolution of 
vorticity perturbations offers further insights into the dynamics uncovered by MFP.
In the initial stages of its evolution, the vorticity perturbations consist of structures inclined against the shear as shown in the inset in Fig.~\ref{fig:PressMFP}(b), where 
the vortical structure marked by the straight-line V1-V2 shows a forward inclination. 
Figures~\ref{fig:PressMFP}(d) and (f) track the movement of points V1 and V2 on this structure as it evolves with time.
Due to its vicinity to the jet core, 
the perturbations 
at point V2 advect faster than V1, 
resulting in a deformation of the vortical structure as it moves downstream. 
Figure~\ref{fig:PressMFP}(d) shows a time instant 
when 
point V2 approaches V1, and the vortical structure is nearly perpendicular to the jet centerline. 
In contrast, 
Fig.~\ref{fig:PressMFP}(f) represents an instance when point V2 convects further downstream of V1. At this instant, 
the vortical structure, which was initially inclined against the shear in Fig.~\ref{fig:PressMFP}(b) is now
inclined with the shear. 
This change in orientation of vortical perturbations 
as they evolve downstream  
indicates the existence of the Orr mechanism discussed in Refs.~\cite{tissot2017wave,tissot2017sensitivity}. 
Additionally, at this instant, the pressure perturbations show the presence of a secondary wavepacket (marked with a rectangle in Fig.~\ref{fig:PressMFP}(e)) which evolves further with time.
In the context of jet instabilities, the evolution of wavepackets due to the Orr mechanism is often referred to as non-modal growth~\cite{tissot2017sensitivity} since it occurs in the convectively stable region of the jet (downstream of the jet exit).
Although the Orr mechanism does not play a significant role in the noise radiation from supersonic jets, 
the presence of this mechanism in the current MFP calculations is encouraging relative to PSE-based models, which do not capture this non-modal growth~\cite{tissot2017sensitivity}.
In fact, the failure of PSE-based models to accurately capture the noise radiation of subsonic jets is often associated with the absence of this mechanism~\cite{schmidt2018spectral}. 
Resolvent-based jet noise models, on the other hand, 
exhibit this mechanism as a response to nonlinear correlated forcing~\cite{schmidt2018spectral}. 
This forcing comprises of inclined vortical structures that are tilted by shear as they are advected downstream analogous to the insets in Figs.~\ref{fig:PressMFP}(b), (d) and (f).
Since MFP does not neglect the nonlinear interaction between various perturbation modes, which appear as forcing terms in resolvent-based models, the existence of the Orr mechanism in the present data is unsurprising.
Future efforts will leverage this capability 
to extend model sound radiation from subsonic jets.

\begin{figure}[!h]
    \centering
    \subfloat[]{\includegraphics[width=0.5\linewidth]{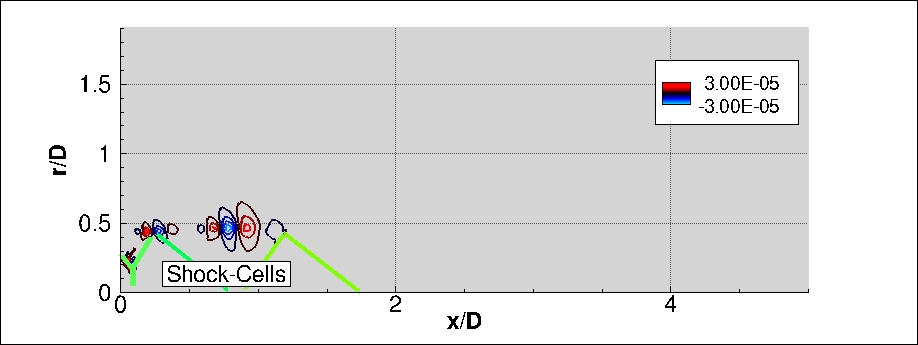}} 
    \subfloat[]{\includegraphics[width=0.5\linewidth]{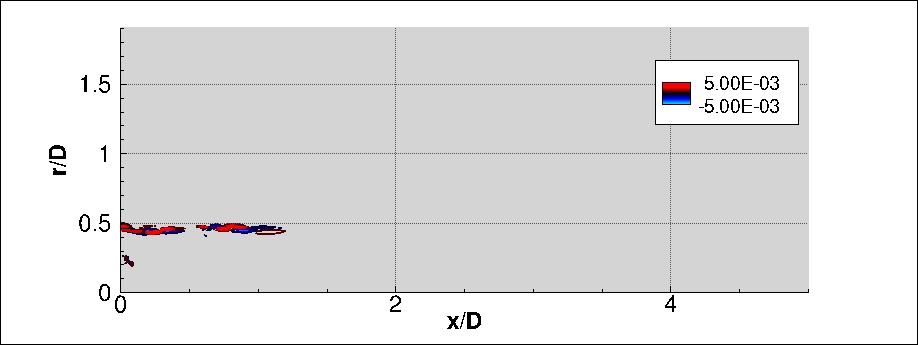}} \\

    \subfloat[]{\includegraphics[width=0.5\linewidth]{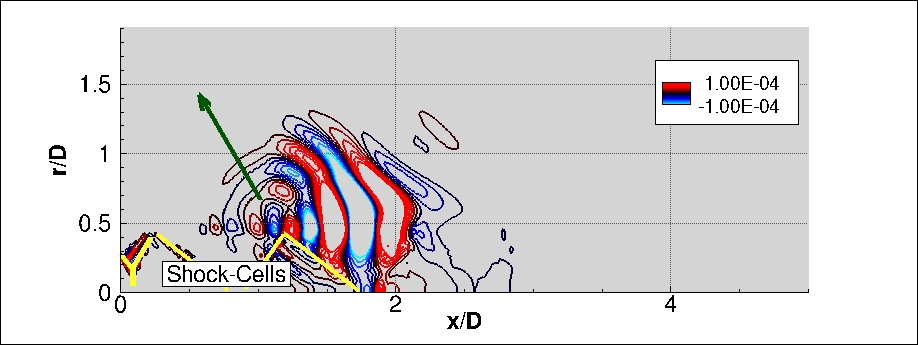}}
     \subfloat[]{\includegraphics[width=0.5\linewidth]{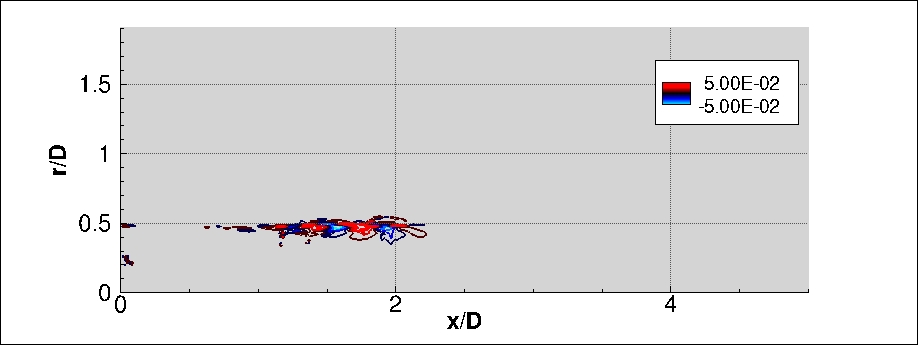}} \\
    \subfloat[]{\includegraphics[width=0.5\linewidth]{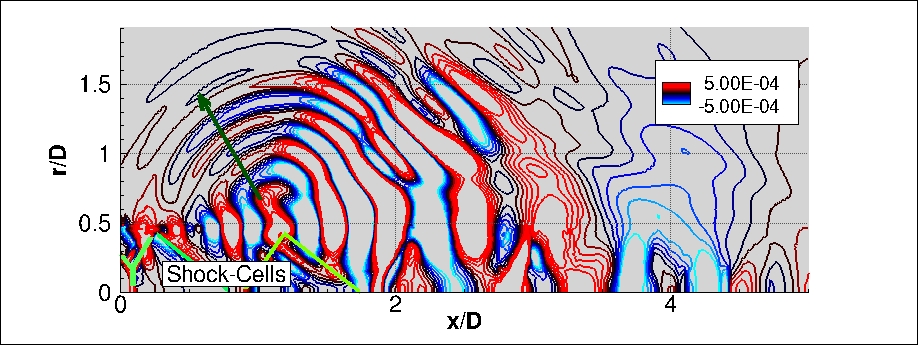}}
      \subfloat[]{\includegraphics[width=0.5\linewidth]{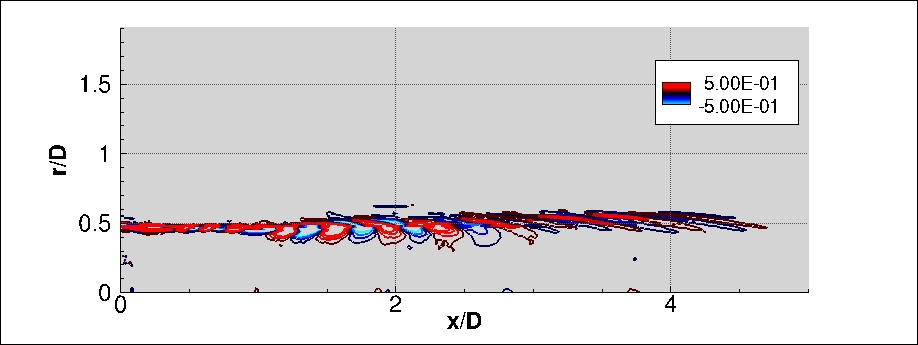}} \\
    \caption{Jet-B: Growth of pressure (left) and vorticity (right) perturbations with time: $t=2.2$ (top), $t=5.2$ (middle), and $t=26.875$ (bottom) . The first few shock-cells are marked for reference.}
    \label{fig:PressMFPB}
\end{figure}
    

The corresponding growth of pressure and vorticity perturbations for Jet-B (Pert04) are shown in Fig.~\ref{fig:PressMFPB}. 
The first few shock-cells immediately downstream of the nozzle exit are marked in Figs.~\ref{fig:PressMFPB}(a), (c) and (e) for reference.
The main downstream radiation signature of the pressure wavepackets in Jet-B is seen to occur at a larger angle from the jet downstream axis when compared to Jet-A. 
This is expected following the wavy-wall analogy~\cite{Tam2009mach}, which models the prominent downstream noise radiation of supersonic jets as supersonically travelling instability waves. 
From the wavy wall analogy, the directivity of the downstream radiation is given by $\cos^{-1}(1/M_c)$, where $M_c$ is the convection Mach number of the instability wave relative to the ambient speed of sound. 
Thus, the shift in downstream directivity away from the jet downstream axis for Jet-B is indicative of a higher convection speed of wavepackets in Jet-B when compared to Jet-A.
Additionally, the interaction of the growing pressure wavepacket with the shock-cells results in the generation of upstream travelling waves resembling BBSAN as indicated by the upward pointing arrows in Figs.~\ref{fig:PressMFPB}(c) and~\ref{fig:PressMFPB}(e). 
Although BBSAN is not the focus of the present study, this observation indicates 
the generality of MFP, which will be leveraged in future efforts.

\subsection{Comparison of Near-field Wavepackets with LES}
\begin{figure}[!h]
    \centering
    \subfloat[]{\includegraphics[width=0.5\linewidth]{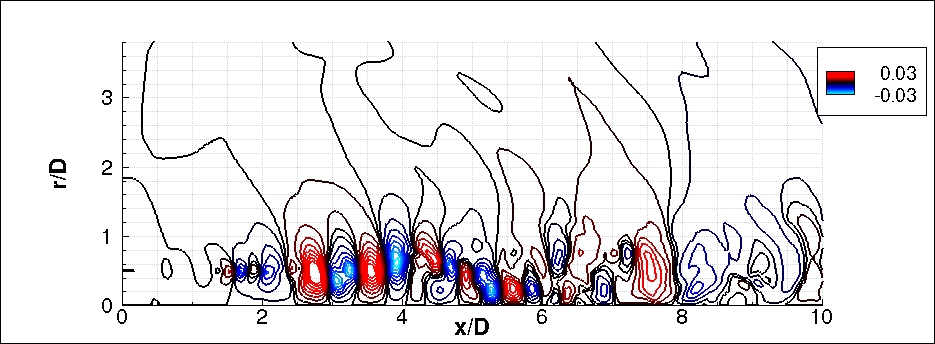}}
    \subfloat[]{\includegraphics[width=0.5\linewidth]{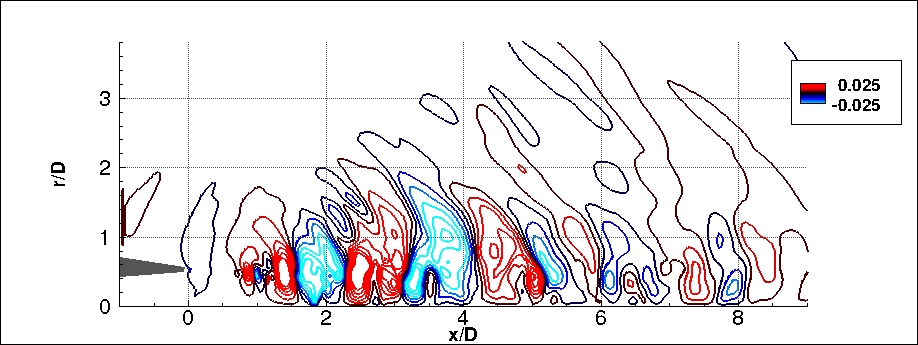}} \\
    \caption{Instantaneous contours of the streamwise component of the acoustic FT Mode from LES of Jet-A (left) and Jet-B (right). }
    \label{fig:FTModes_LES}
\end{figure}

The comparison of MFP-based predictions of jet noise with the original LES is performed through the acoustic component discussed in $\S$\ref{section:Doak}. 
Figure~\ref{fig:FTModes_LES} 
shows the streamwise component of the acoustic component from the LES fluctuation data 
for the two jets 
at a representative time-step. 
Similar observations can be made on the downstream directivity for the two jets as in Figs.~\ref{fig:PressMFP} and~\ref{fig:PressMFPB}, where 
the main radiation signature of Jet-B, due to its heated nature, propagates at a larger downstream angle when compared to Jet-A.
For both cases, the 
acoustic mode is comprised of a coherent wavepacket structure representative of classical instability waves~\cite{tam1980radiation,tam1984sound,tam1984sound2} used to model the sound generation from jets. 

\begin{figure}[!h]
    \centering
        \subfloat[]{\includegraphics[width=0.5\linewidth]{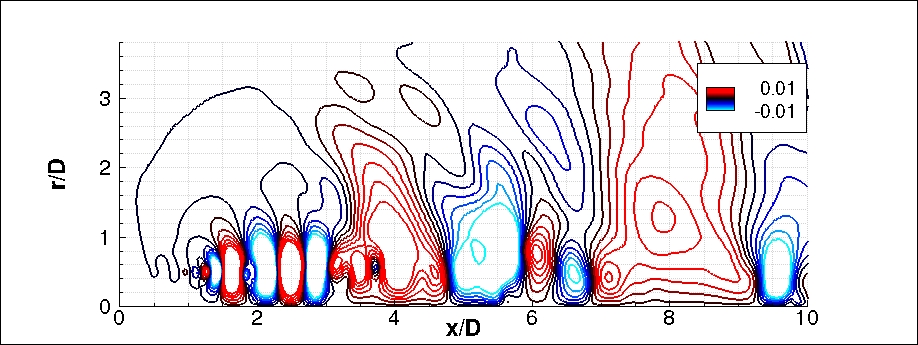}}
    \subfloat[]{\includegraphics[width=0.5\linewidth]{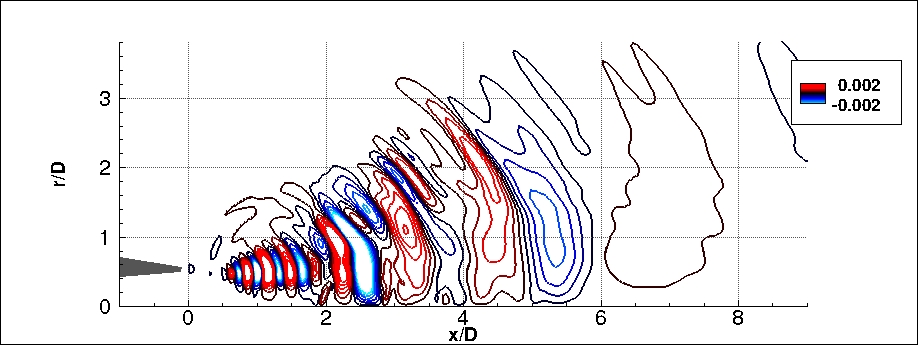}}
    \caption{Instantaneous contours of the streamwise component of the acoustic FT perturbations from MFP of Jet-A (left) and Jet-B (right). }
    \label{fig:FTModes_MFP}
\end{figure}

Despite the similarity to pressure fluctuations  upstream of the core collapse region, it has been shown elsewhere for Jet-A~\cite{unnikrishnan2019acoustically} and Jet-B~\cite{prasad2019modal,prasad2020study}, that the acoustic component exhibits superior spatio-temporal coherence and radiative efficiency for the LES data.
These observations also hold when using perturbations 
propagating on
the mean flow.
Figure~\ref{fig:FTModes_MFP} displays the acoustic component of the perturbation field for each of the two jets after it has filled the domain. 
The acoustic components of the LES and MFP are not expected to be the same, since the former contains all resolved scales in the flow, while the latter represents propagation on the mean. 
Nonetheless, the more compact nature of the acoustic perturbation is easily evident from a comparision between Fig.~\ref{fig:FTModes_MFP} and the pressure fields of Figs.~\ref{fig:PressMFP} and~\ref{fig:PressMFPB}.
We thus use $-\frac{\partial \psi_a}{\partial x}$ as a measure of the near acoustic field for the remainder of this investigation.

The ability of the acoustic component ($-\frac{\partial \psi_a}{\partial x}$) from the MFP-propagated perturbations to accurately capture the noise generation trends is now examined by comparing with the corresponding LES results.
For this, we make use of data-driven DMD~\cite{schmid2010dynamic}, which processes a series of snapshots to identify spatial modes that optimally describe the evolution of the flow from one time instant to another.
Briefly, with $N$ snapshots spaced at constant sampling time-step-sizes and assembled in the subspace $\textbf{U}$, DMD provides the following representation 
\begin{equation} \label{eqn:DMD}
    \textbf{U}(\textbf{x},t)= \sum_{n=1}^N a_n \exp{(\lambda_n t)}\Phi_n(\textbf{x})
\end{equation}
where $a_n$, $\Phi_n(\textbf{x})$ and $\lambda_n$ represent the amplitude, spatial structure and the complex frequency of the $n$-th mode respectively. 
The complex frequency can be further split into $\lambda_n=\sigma + i\omega$, where, $\sigma$ and $\omega$ represent the growth rate and the frequency of the mode respectively. 
Although DMD modes are associated with their specific frequencies, they may also be organized by their amplitudes as defined in Ref.~\cite{jovanovic2014sparsity}.

When used with MFP, DMD provides a robust and economical option to calculate the growth rates of the unstable modes for a wide range of both internal as well as external flows \cite{ranjan2020robust}. 
For the statistically stationary LES data, with the exception of a small number of outliers, the DMD modes are tightly clustered along a unit circle in the Ritz space and therefore have nearly zero growth/decay rates.
The perturbation growths obtained from MFP are useful however, especially in the context of control ($\S$\ref{section:Control}). 
The comparison of the LES and MFP results is thus performed by considering DMD modes with the highest amplitudes from the LES.
Three ranges are selected corresponding to  the low- $(0.2 \leq St \leq 0.45)$, mid- $(0.45 < St \leq 0.75)$ and high-frequency $(St > 0.75)$ regimes.
In these, the highest amplitude LES fluctuation DMD modes are compared with the corresponding MFP perturbation DMD modes at the same frequencies.

\begin{figure}[!h]
    \centering
    \subfloat[$St=0.35$]{\includegraphics[width=0.5\textwidth]{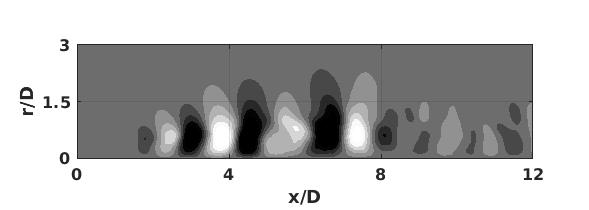}} 
    \subfloat[$St=0.35$]{\includegraphics[width=0.5\textwidth]{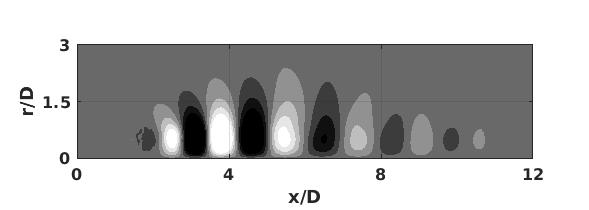}}  \\
     \subfloat[$St=0.56$]{\includegraphics[width=0.5\textwidth]{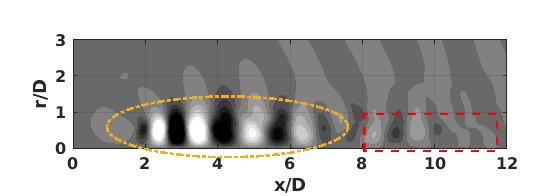}} 
        \subfloat[$St=0.56$]{\includegraphics[width=0.5\textwidth]{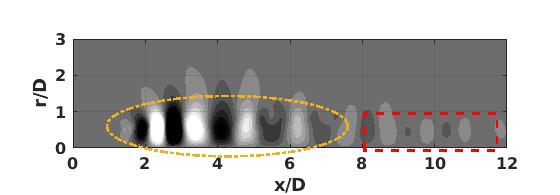}} \\
    \subfloat[$St=0.90$]{\includegraphics[width=0.5\textwidth]{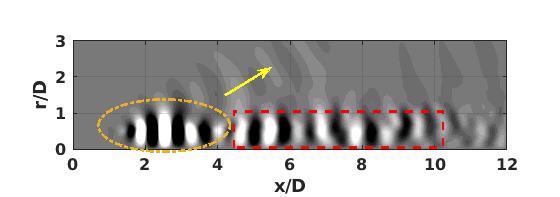}}
        \subfloat[$St=0.90$]{\includegraphics[width=0.5\textwidth]{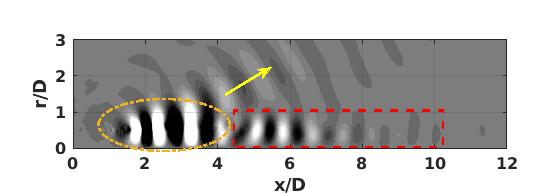}} \\
    \caption{Jet-A: Comparison of the most dominant acoustic DMD modes from LES (left) with MFP (right) at different \emph{St} values corresponding to low, mid and high frequency regimes (Pert01).}
    \label{fig:NearfieldWavepackets}
\end{figure}

\begin{figure}[!h]
    \centering
    \subfloat[$St=0.35$]{\includegraphics[width=0.5\linewidth]{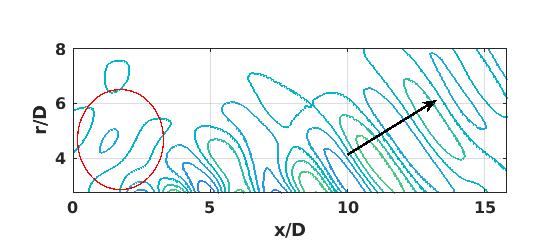}}
    \subfloat[$St=0.35$]{\includegraphics[width=0.5\linewidth]{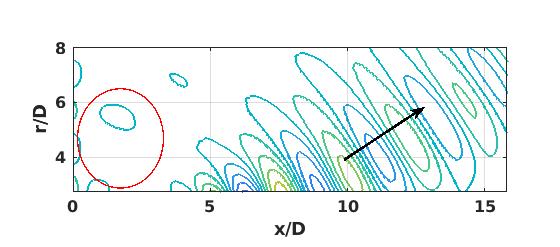}}\\
    \subfloat[$St=0.56$]{\includegraphics[width=0.5\linewidth]{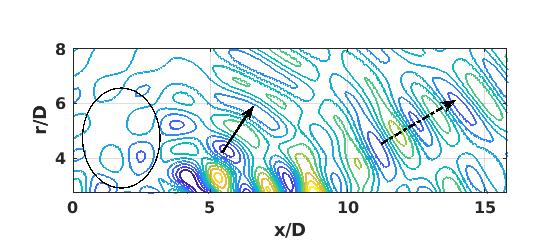}}
    \subfloat[$St=0.56$]{\includegraphics[width=0.5\linewidth]{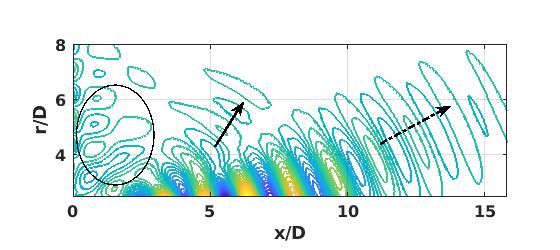}}\\
    \subfloat[$St=0.90$]{\includegraphics[width=0.5\linewidth]{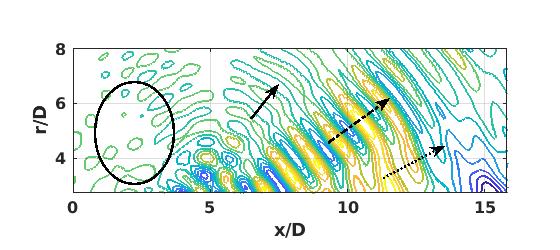}}
    \subfloat[$St=0.90$]{\includegraphics[width=0.5\linewidth]{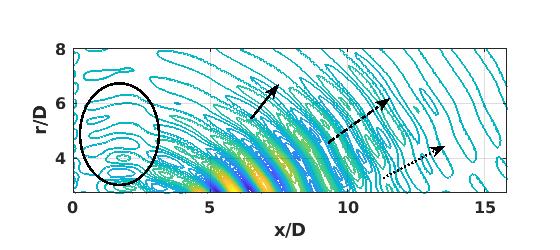}}\\
    
        \caption{Jet-A: Comparison of noise radiation patterns from the wavepackets shown in Fig.~\ref{fig:NearfieldWavepackets} from LES (left) and MFP (right). The radiation patterns are generated by plotting the $-\frac{\partial \psi_a}{\partial x}$ DMD modes in the domain $0 \leq x/D \leq 16$ and $3 \leq r/D \leq 8$.}
    \label{fig:Radiation}
\end{figure}

Figure~\ref{fig:NearfieldWavepackets} shows a comparison of the  near-field $-\frac{\partial \psi_a}{\partial x}$ DMD modes in the form of wavepackets obtained from both LES and MFP data with increasing frequency for Jet-A. 
Figure~\ref{fig:Radiation} on the other hand, shows a qualitative picture of the acoustic radiation due to each of these wavepackets.
Both these figures are obtained by plotting the $-\frac{\partial \psi_a}{\partial x}$ DMD modes in different sub-domains of the flow-field.
Several important observations make evident that the mechanisms identified in the LES-based  $-\frac{\partial \psi_a}{\partial x}$ modes are active in the MFP and provide the proper basis for the prediction technique. 
In the low frequency regime ($St=0.35$), the wavepacket represented by the $-\frac{\partial \psi_a}{\partial x}$ LES DMD mode (Fig.~\ref{fig:NearfieldWavepackets}(a)) displays a single coherent streamwise structure, which is linked to the K-H instability of the jet shear layer~\cite{schmidt2018spectral}. 
This results in a corresponding prominent downstream radiation pattern as shown with the arrow in Fig.~\ref{fig:Radiation}(a).
This general wavepacket structure shape, its streamwise extent and radiation pattern are properly reproduced by the linear technique
(Figs.~\ref{fig:NearfieldWavepackets}(b) and~\ref{fig:Radiation}(b)).
This is expected 
because the white-noise nature of the initial perturbation in the current procedure is similar to K-H forcing in previous studies of jet noise with input-output systems as discussed previously. 

At the next higher frequency, $St=0.56$, focusing first on the LES  (Fig.~\ref{fig:NearfieldWavepackets}(c)), the overall wavepacket in the corresponding DMD mode is more complicated with the addition of a secondary streamwise coherent structure downstream of the core collapse.
This secondary wavepacket structure appears related to the Orr mechanism discussed previously.
The primary K-H wavepacket persists at these frequencies (highlighted with a yellow oval) but becomes more confined to the jet exit \textit{i.e.,}  the axial envelope of the wavepacket is shortened.
These features of the LES-derived wavepacket structures at this frequency 
are also evident in the corresponding acoustic perturbations from MFP as shown in Fig.~\ref{fig:NearfieldWavepackets}(d), which also shows the primary K-H wavepacket as well as the secondary downstream streamwise coherent structure. 
The radiation patterns at $St=0.56$  are shown in Fig.~\ref{fig:Radiation}(c) for LES and Fig.~\ref{fig:Radiation}(d) for the MFP.
The increased complexity of the wavepacket is manifested in the existence of more than one acoustic beam; at this frequency, two prominent directions are observed as marked by arrows.

At the highest frequency considered, $St=0.9$, the observations on trends and comparisons between LES and MFP wavepackets (Figs.~\ref{fig:NearfieldWavepackets}(e) and~(f), respectively) as well as their radiation patterns (Figs.~\ref{fig:Radiation}(e) and~(f), respectively) remain valid.
Specifically, the primary wavepacket becomes shorter, while the secondary one lengthens.
The most prominent radiation 
patterns appear in more directions as marked by arrows.

Finally, the upstream radiation footprint at all these frequencies (shown by circles in Fig.~\ref{fig:Radiation}) are also exhibited by the perturbation DMD modes.
These observations demonstrate that the primary noise generation mechanisms existing in the full LES are  captured by the MFP.
Furthermore, the existence of sideline radiation signatures in the perturbation field suggests that MFP can used to model the contribution of large-scale structures to the sideline noise which may have feedback control benefits with wing mounted~sensors~\cite{cheng2021two}.

\begin{figure}[!h]
    \centering
    \subfloat[$St=0.30$]{\includegraphics[width=0.5\textwidth]{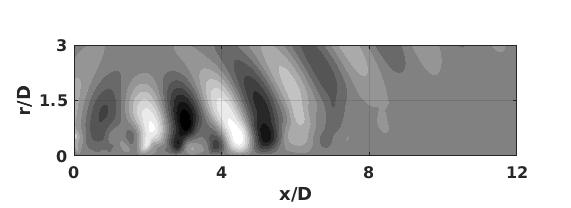}} 
    \subfloat[$St=0.30$]{\includegraphics[width=0.5\textwidth]{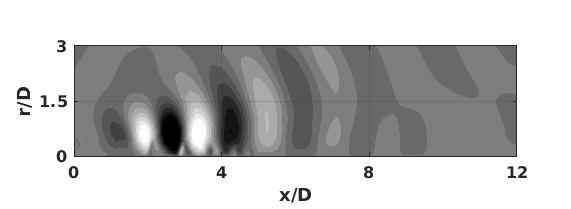}}  \\
     \subfloat[$St=0.48$]{\includegraphics[width=0.5\textwidth]{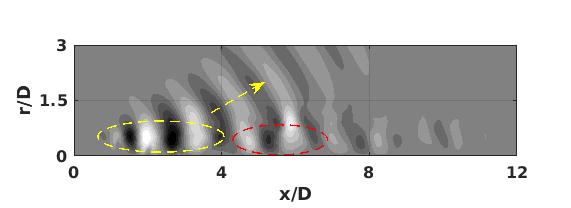}} 
        \subfloat[$St=0.48$]{\includegraphics[width=0.5\textwidth]{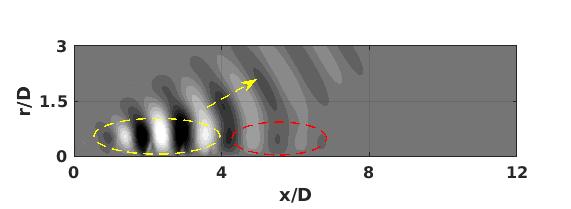}} \\
    \subfloat[$St=0.93$]{\includegraphics[width=0.5\textwidth]{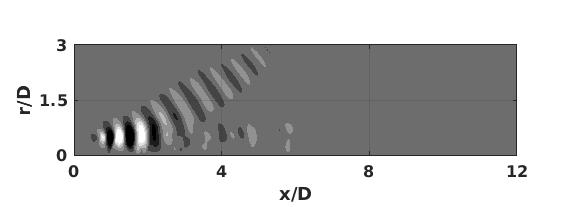}}
       \subfloat[$St=0.93$]{\includegraphics[width=0.5\textwidth]{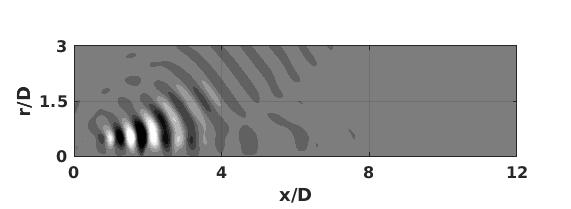}} \\
    \caption{Jet-B (Pert04): Comparison of the most dominant acoustic DMD modes from LES (left) with MFP (right) at different \emph{St} values corresponding to low, mid and high frequency regimes.}
    \label{fig:NearfieldWavepacketsGE404}
\end{figure}

Figure~\ref{fig:NearfieldWavepacketsGE404} shows a comparison of the near-field $-\frac{\partial \psi_a}{\partial x}$ DMD modes for Jet-B for the test case Pert04. 
The radiation signature due to each of these wavepackets is also visible in this figure, and are thus  not plotted separately.
As expected, due to the heated nature of the jet, the dominant DMD modes at each frequency have a single streamwise structure with a strong downstream radiation signature, analogous to a supersonically convecting wavy wall~\cite{Tam2009mach}.
This agrees with previous observations based on Spectral Proper Orthogonal Decomposition (SPOD) of the acoustic mode in Ref.~\cite{prasad2020study} and follows from the fact that SPOD modes can be interpreted as optimally averaged DMD modes from an ensemble DMD problem~\cite{towne2018spectral}.
Additionally, similar to Jet-A, with an increase in frequency, the wavelength of the radiation is shorter and the streamwise extent of the Mach wave emitting region is confined closer to the nozzle exit. 
This is also consistent with the noise source characteristics of jets, where the higher frequencies are radiated
from closer to the nozzle exit.
The K-H type wavepacket, its subsequent radiation and trends observed with increasing frequency are accurately captured by the current procedure despite the existence of strong shocks in the jet plume; this demonstrates its ability to model sound from practical shock-containing military configurations.

\subsection{Spectral Comparison}
\begin{figure}[!h]
    \centering
    \subfloat[]{\includegraphics[width=0.33\textwidth]{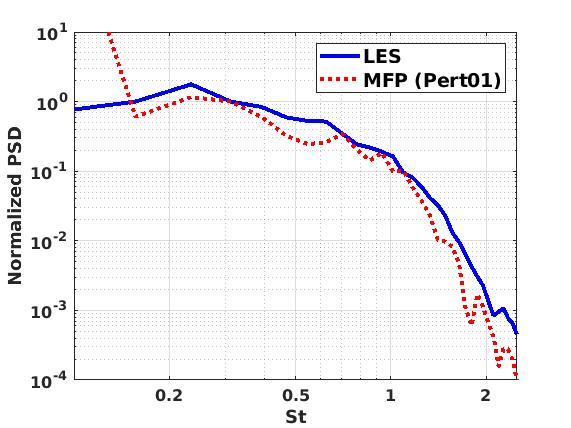}}
        \subfloat[]{\includegraphics[width=0.33\textwidth]{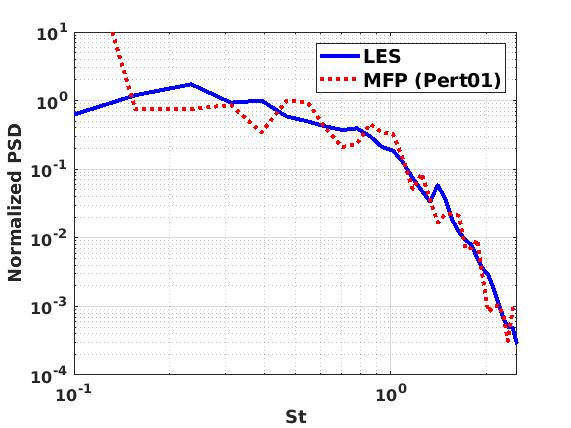}} 
            \subfloat[]{\includegraphics[width=0.33\textwidth]{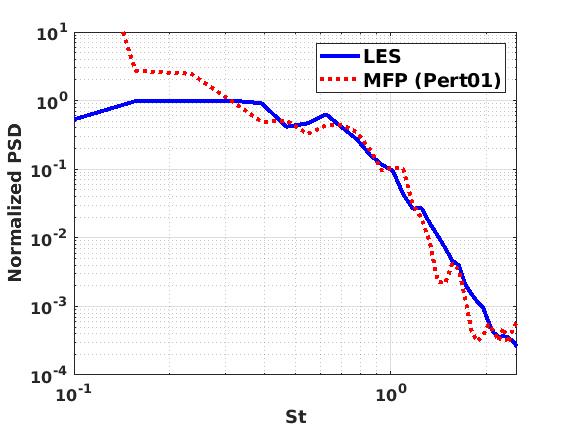}}
    \caption{Jet-A: Normalized power spectral density calculated from MFP compared with LES at (a) $\theta=30^\circ$, (b) $\theta=45^\circ$, and (c) $\theta=60^\circ$ at $r/D=8$.}
    \label{fig:PSD_normA}
\end{figure}

\begin{figure}[!h]
    \centering
    \subfloat[]{\includegraphics[width=0.33\textwidth]{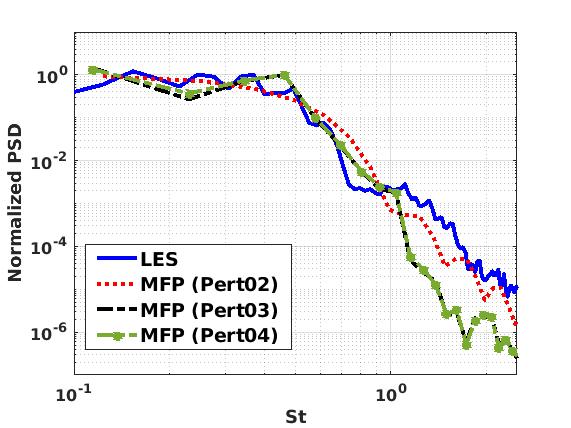}}
        \subfloat[]{\includegraphics[width=0.33\textwidth]{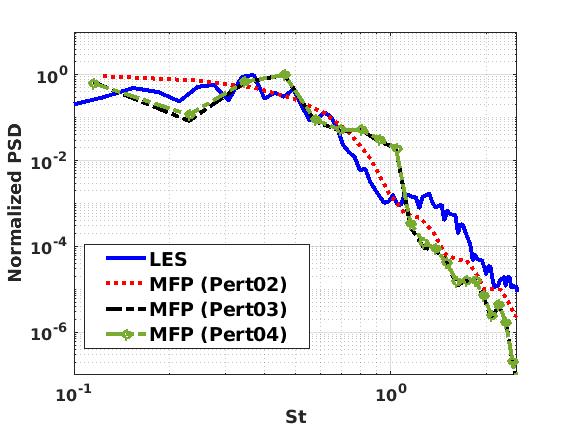}} 
            \subfloat[]{\includegraphics[width=0.33\textwidth]{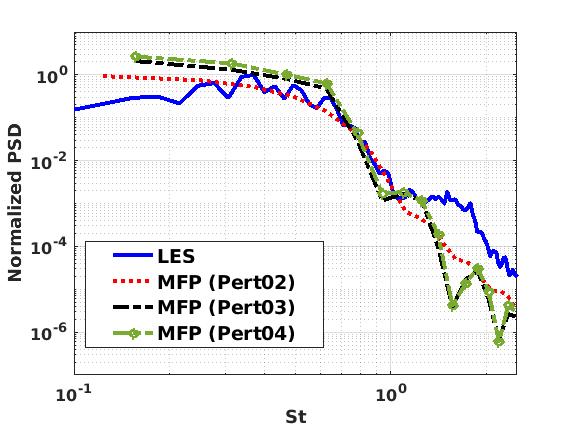}}
    \caption{Jet-B: Normalized power spectral density calculated from MFP compared with LES at (a) $\theta=40^\circ$, (b) $\theta=50^\circ$, and (c) $\theta=60^\circ$ at $r/D=8$.}
    \label{fig:PSD_normB}
\end{figure}

Near-field spectra are now computed to provide a more quantitative comparison for all the test cases documented previously in Table~\ref{tab:MFPCases}.
For this, we select a location at $8D$ along several angles $\theta$ measured from the axis -- these locations are in the near acoustic field where 
the LES and MFP meshes are sufficiently fine to capture the unsteadiness.
The hydrodynamic content at these locations is negligible and therefore  $-\partial \psi'_a/\partial x$ is a scaled surrogate for pressure fluctuations~\cite{unnikrishnan2016acoustic,prasad2020study}.
Figures~\ref{fig:PSD_normA} and~\ref{fig:PSD_normB} display the normalized power spectral density (PSD) of $p'$, from LES and MFP along three angles where the prominent noise field is obtained for Jet-A and Jet-B, respectively.
The PSD is normalized by the maximum value at \emph{St} $\geq 0.2$.
The general shape of the reconstructed PSD from MFP matches that of the LES for most of the \emph{St} values for Jet-A and for $0.2 \leq St \leq 1.0$ for Jet-B.
This is consistent with the favorable qualitative comparisons of the DMD modes at these frequencies shown previously.
At the higher angles, MFP overpredicts the PSD at low frequencies, with the disagreement extending to larger \emph{St} with increasing polar angle from the jet downstream axis.
This disagreement is similar to that observed with other stability-based approaches whose comparisions break down at very low frequencies in more striking ways than with the present method.
One possible reason is associated with the relative unimportance of the K-H instability at these low \emph{St} values in the jet flow-field as discussed in Ref.~\cite{schmidt2018spectral}.
When this range of frequencies is of interest, an alternative approach is to replace the white-noise impulse with a continuous forcing that explicitly introduces the desired ranges.

Another pertinent observation is that the general trends  for Jet-B are insensitive to the perturbation magnitude and the initial forcing location as long as the initial forcing encompasses the thin shear layer near the nozzle exit. 
Furthermore, the recovered trends for the test cases Pert03 and Pert04, which
share the same forcing location but are computed on different meshes with $\epsilon_0$ values that are an order of magnitude apart, are almost identical.
This not only demonstrates the linearity of the present approach for these typical parameters but also
provides evidence that the mesh requirements for MFP are considerably less stringent that those of the LES, as the same spectra are recovered on one-fourth of the original LES mesh. 


\section{Effect of Control Frequency \label{section:Control}}
The perturbations obtained from the current time-domain procedure, with a suitably chosen forcing, contain information on growth rates that can be leveraged to inform the response to active control through excitation.
All components of the current approach, MFP, sifting and DMD contribute to this endeavor.
Although there is no restriction of the approach on properties of the basic state, here we perform the analysis on Jet-A, because extensive experimental results using Localized Arc Filament Plasma Actuators (LAFPAs) and corresponding LES have successfully examined and characterized the effect of actuator frequency on noise~\cite{samimy2007active,gaitonde2011coherent}.
This facilitates a clear method to evaluate the trend predictions.

Although LAFPAs can be employed for diagnostic purposes by phase-locking instabilities, they have been shown to exhibit control authority over different high-speed flow configurations.
A comprehensive review of these actuators for high-speed flow control can be found in Ref.~\cite{samimy2019reinventing}.
Briefly, in the context of jet noise, LAFPAs are designed to introduce perturbations in the jet shear layer by producing a high-temperature arc discharge near the nozzle exit. 
Since the K-H instability of the jet shear layer is sensitive to perturbations close to the nozzle exit, mixing enhancements can be obtained by pulsing these actuators at specific frequencies and wavenumbers, resulting in noise mitigation.

\begin{figure}[!h]
    \centering
    \includegraphics[width=0.8\textwidth]{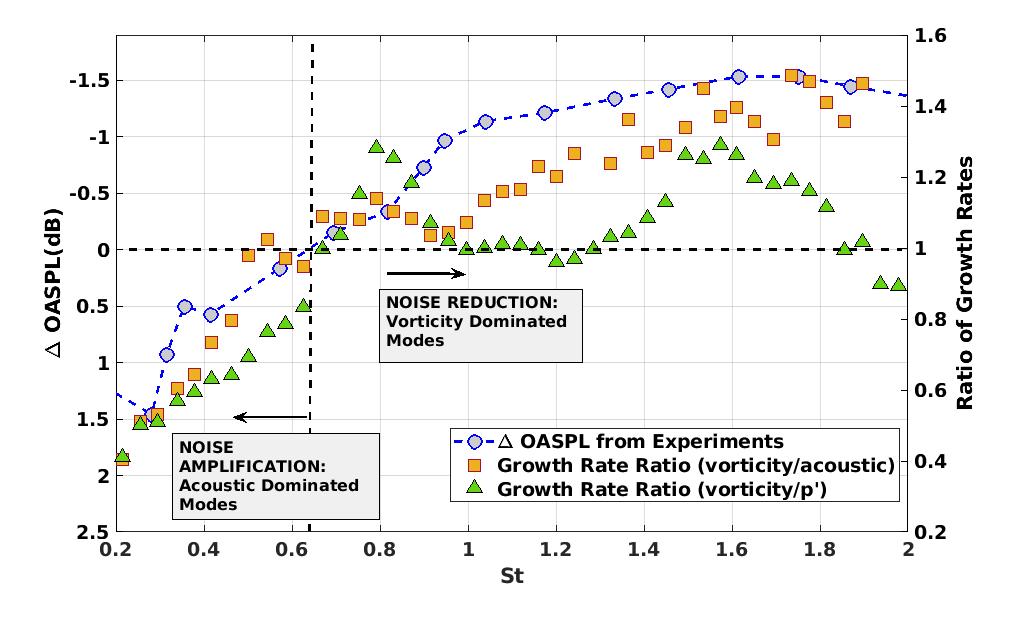}
    \caption{Noise reduction using LAFPA actuators for Jet-A~\cite{samimy2010acoustic} marked with circles, and ratio of growth rates for vorticity and $\bm {-\frac{\partial \psi_a}{\partial x}}$ perturbations from MFP marked with squares. Triangles denote the ratio of growth rates for vorticity and pressure perturbations from MFP.}
    \label{fig:Control}
\end{figure}

Figure~\ref{fig:Control} shows the experimental measurements in the form of differences in Overall Sound Pressure Levels ($\Delta$OASPL, shown with blue circles), relative to the no control case, in the direction of peak noise radiation $(\theta=30^\circ)$ as a function of frequency~\cite{samimy2010acoustic}. 
These results correspond to an in-phase axisymmetric 
actuation mode in the original experiment. 
Note that the $y-$axis is flipped such that the $\Delta$OASPL values above the horizontal dashed line represent noise reduction and vice-versa. 
It is evident that pulsing the actuators at forcing frequencies near the jet preferred-mode $(St \approx 0.3)$ results in the maximum amplification of downstream noise. 
The coherent structures are comprised of large vortex rings with braids~\cite{gaitonde2011coherent}.
In contrast, noise mitigation is achieved at higher forcing frequencies with a maximum $\Delta$OASPL of the order of $-1.5dB$ at $St \approx 1.6$. 
At these frequencies, mixing is enhanced without the formation of large coherent structures.

Since MFP can accurately capture the spatio-temporal signature of the K-H instability, it can be tailored to predict the noise mitigation/enhancement resulting from  actuation.
This is achieved by restricting the initial randomized perturbation given by Eq.~(\ref{eqn:forcing}) with $\epsilon_0=10^{-5}$ 
to only a small region at the nozzle sleeve at the actuator location as in the simulations of Ref.~\cite{gaitonde2012analysis}. 
The growth of perturbations from the initial forcing is analyzed using the same steps as before.
A total of $1{,}000$ snapshots are sampled at $\Delta t=0.025$; 
this combination of $\epsilon_0$, $\Delta t$ and number of snapshots is found to be appropriate to maintain a linear growth of perturbations based on simple proportionality tests between input and output (not shown).
Due to the implicit linearization procedure of MFP, the contribution from second-order terms in Eq.~(\ref{eqn:MFP_expanded}) can become significant as the perturbation field grows with time.
Non-linear perturbation evolution, although beneficial in some cases (for example, to explore intermittency~\cite{prasad2021extraction}), prohibits an accurate assessment of growth rates from DMD which has control implications.


As mentioned previously in $\S$\ref{section:Intro}, most noise control methods seek to break up the noise-efficient large-scale structures in the jet shear layer by enhancing mixing. 
Therefore, the actuation frequencies associated with noise reduction must be vorticity dominated. 
In the current framework, this reduces to a problem of identifying the frequencies of the DMD modes that have a high growth rate of vorticity. 
At the same time, these frequencies should have a lower growth rate of the $-\frac{\partial \psi_a}{\partial x}$ perturbations so that the mixing dominates the generation of new acoustic waves due to the actuation. 
The growth rates are obtained as the real part of the complex frequency $\lambda_n$ in Eq.~(\ref{eqn:DMD}) by performing DMD separately on both vorticity as well as $-\frac{\partial \psi_a}{\partial x}$ perturbation snapshots.
The frequencies resolved by DMD (and hence their growth rates) are directly related to the total number of snapshots and their sampling rate. 
Therefore, while subjecting the snapshots to DMD, the subspace size was varied from $200$ to $1{,}000$ snapshots starting from the initial perturbation at $t=0$ with different sampling frequencies to test for convergence of growth rates following the guidelines listed in Ref.~\cite{ranjan2020robust}.
Special care was taken to confirm that the conclusions drawn from 
this section are insensitive to these choices.

A suitable measure to examine the relative growth of the vorticity component to the MPT acoustic variable is to consider the ratio of the real part of the Ritz values.
A map of the ratio of growth rates of vorticity and $-\frac{\partial \psi_a}{\partial x}$ perturbations as a function of \emph{St} provided in Fig.~\ref{fig:Control} (squares) using the right $y-$axis.
Note that the horizontal line representing $\Delta$OASPL $=0$ for the experimental data on the left~$y-$axis corresponds to a growth rate ratio of one on the right~$y-$axis. 
This horizontal line is used to demarcate the ``vorticity-dominated" and ``acoustic-dominated" DMD modes as marked in the Fig.~\ref{fig:Control}. 
The growth rates are restricted to $St \geq 0.2$, since the noise predictions from MFP deviate from the LES results at very low frequencies as seen previously in Fig.~\ref{fig:PSD_normA} for Jet-A. 
In the limit $St\rightarrow 0$, the growth rates of both quantities of interest, vorticity and the MPT acoustic variable, become small, and their ratio becomes ill-behaved.

At lower frequencies $St \leq 0.5$, $-\frac{\partial \psi_a}{\partial x}$ is associated with very high growth rates when compared to vorticity. 
This suggests that actuation at these frequencies results in an increase in acoustic disturbances in the shear layer with lower mixing. 
This is confirmed by the experimental observations indicating that actuation in this frequency band results in an increase in noise radiated from the jet. 
In contrast, at higher frequencies the growth of vorticity in the shear layer increases relative to that of the acoustic perturbations. 
Thus, actuation at these higher frequencies results in an increased rate of mixing in the shear layer. 
However, acoustic disturbances generated due to actuation at this frequency do not grow as significantly when compared with mixing. 
This implies any actuation in the frequency range $0.65 \leq St \leq 2$ should result in a decrease in noise radiated by the jet. 
This is in remarkable agreement with the experimental measurements and supports the observation that the identification of the ``vorticity dominated" frequency range can inform noise control technologies for nozzle designs. 

For completeness, Fig.~\ref{fig:Control} also shows the ratio of vorticity to pressure perturbation growth rates as a function of \emph{St} value (marked with triangles).
As expected, the use of pressure instead of $-\frac{\partial \psi_a}{\partial x}$  in the denominator does not provide such a clear cut demarcation between the vorticity and acoustic dominated modes, since pressure perturbations also contain contributions from the non-radiating component. 
Hence, the growth rates of pressure perturbations cannot be interpreted as an accurate measure of sound generation.
The sifting of acoustic perturbations is therefore a key component if the present technique is to be applied for control trend analysis. 
If the method is used for noise prediction only, this sifting, although highly advantageous, is not entirely necessary as away from the jet shear layer, the pressure perturbations and acoustic component are related through a simple scaling.

Finally, it must be noted, that the success of the linear method in predicting noise generation and control trends with actuator frequency does not indicate that the noise generation from the jet is a linear process. 
The underlying turbulent mean basic state is necessary to calculate the growth of perturbations and ensures that elements of non-linearity that play an important role in sound generation are retained.


\section{Conclusion \label{conclusion}}

This investigation presents an implicit time-domain approach to model the noise radiation from on- and off-design supersonic jets as well as to estimate noise control trends with actuator frequencies.  
This is achieved in three steps.
The first involves adding an easily obtained body-force term to an existing Navier-Stokes (NS) code; this converts the native NS solver into an implicit linearized NS solver (LNSE) that can compute the linear growth of perturbations about the long-time mean of a turbulent flow-field.
The implicit linearization avoids additional constraints on configuration and facilitates a straightforward extension to off-design operation conditions such as those found in military-style jets, which are otherwise challenging to model using existing techniques.

To test the robustness of the present approach, two LES databases are employed as truth models. 
These represent an ideally-expanded Mach~1.3 jet from a thin sleeve nozzle using an in-house code and a heated over-expanded Mach~1.36 jet from a faceted military-style nozzle using a widely used commercial code.
The robustness of the body-force specification allows the use of the same LNSE solver to compute the evolution of linear perturbations about time-averaged basic states for both test cases.
The forcing employed consists of white-noise impulse pressure forcing in the jet shear layer.
Mean flow gradients then induce perturbations in all other primitive quantities.
The acoustic component is obtained by sifting the resulting perturbation field using Doak's momentum potential theory. 
By utilizing data-driven Dynamic Mode Decomposition (DMD) to identify growing spatial modes, it is demonstrated that the primary noise radiation mechanisms existing in the LES fluctuations are reproduced by the acoustic perturbations, irrespective of the operating conditions.
Moreover, away from the jet shear layer, the noise spectra from the LNSE matches the desired variations obtained from the LES for a range of frequencies of interest at a fraction of the cost of the original LES, highlighting the cost-effective nature of the present approach.

For control trends, 
it is postulated that by localizing the initial forcing to actuator locations as in the experiment, the identification of the ``vorticity-dominated" frequency range, where the growth rate of vorticity perturbations dominates the growth rate of acoustic perturbations represents the desired actuator frequencies for noise control.
The results are in excellent agreement with previously reported experimental data and shows the potential of the approach 
to model sound radiation and inform control techniques in a robust and computationally efficient manner. 
Athough the present study is focused on axisymmetric jet configurations due to their well-known dynamics, this approach can be easily extended to non-axisymmetric jets at off-design operating conditions by leveraging the capabilities of the native NS solver which is often a limitation of most existing jet noise models.
Current efforts are focused on exploring the use of the method for rectangular jets of different aspect ratios.

\section*{Acknowledgements}
This work was performed in part under the sponsorship of the Office of Naval Research with Dr. S. Martens serving as Project Monitor (Technical points of contact: J. Spyropoulos and R. Powers).
The views and conclusions contained herein are those of the authors and do not represent the opinion of the Office of Naval Research or the U.S. government. The authors are also grateful to Dr. Rajesh Ranjan
for insightful discussions on the implementation of MFP to the LES data.

\bibliography{Ref}

\begin{thebibliography}{10}
\expandafter\ifx\csname url\endcsname\relax
  \def\url#1{\texttt{#1}}\fi
\expandafter\ifx\csname urlprefix\endcsname\relax\def\urlprefix{URL }\fi
\expandafter\ifx\csname href\endcsname\relax
  \def\href#1#2{#2} \def\path#1{#1}\fi

\bibitem{haines2001chronic}
M.~M. Haines, S.~A. Stansfeld, R.~S. Job, B.~Berglund, J.~Head, Chronic
  aircraft noise exposure, stress responses, mental health and cognitive
  performance in school children, Psychological medicine 31~(2) (2001)
  265--277.
\newblock \href {https://doi.org/https://doi.org/10.1017/S0033291701003282}
  {\path{doi:https://doi.org/10.1017/S0033291701003282}}.

\bibitem{schmidt2013effect}
F.~P. Schmidt, M.~Basner, G.~Kr{\"o}ger, S.~Weck, B.~Schnorbus, A.~Muttray,
  M.~Sariyar, H.~Binder, T.~Gori, A.~Warnholtz, et~al., Effect of nighttime
  aircraft noise exposure on endothelial function and stress hormone release in
  healthy adults, European heart journal 34~(45) (2013) 3508--3514.
\newblock \href {https://doi.org/https://doi.org/10.1093/eurheartj/eht269}
  {\path{doi:https://doi.org/10.1093/eurheartj/eht269}}.

\bibitem{martens2010practical}
S.~Martens, J.~T. Spyropoulos, Practical jet noise reduction for tactical
  aircraft, in: ASME Turbo Expo 2010: Power for Land, Sea, and Air, American
  Society of Mechanical Engineers Digital Collection, 2010, pp. 389--399.
\newblock \href {https://doi.org/https://doi.org/10.1115/GT2010-23699}
  {\path{doi:https://doi.org/10.1115/GT2010-23699}}.

\bibitem{tam1971directional}
C.~K. Tam, Directional acoustic radiation from a supersonic jet generated by
  shear layer instability, Journal of Fluid Mechanics 46~(4) (1971) 757--768.
\newblock \href {https://doi.org/https://doi.org/10.1017/S0022112071000831}
  {\path{doi:https://doi.org/10.1017/S0022112071000831}}.

\bibitem{sarohia1978experimental}
V.~Sarohia, P.~Massier, Experimental results of large-scale structures in jet
  flows and their relation to jet noise production, AIAA Journal 16~(8) (1978)
  831--835.
\newblock \href {https://doi.org/https://doi.org/10.2514/3.60969}
  {\path{doi:https://doi.org/10.2514/3.60969}}.

\bibitem{troutt1982experiments}
T.~Troutt, D.~McLaughlin, Experiments on the flow and acoustic properties of a
  moderate-reynolds-number supersonic jet, Journal of Fluid Mechanics 116
  (1982) 123--156.
\newblock \href {https://doi.org/https://doi.org/10.1017/S0022112082000408}
  {\path{doi:https://doi.org/10.1017/S0022112082000408}}.

\bibitem{jordan2013wave}
P.~Jordan, T.~Colonius, Wave packets and turbulent jet noise, Annual review of
  fluid mechanics 45 (2013) 173--195.
\newblock \href
  {https://doi.org/https://doi.org/10.1146/annurev-fluid-011212-140756}
  {\path{doi:https://doi.org/10.1146/annurev-fluid-011212-140756}}.

\bibitem{gudmundsson2011instability}
K.~Gudmundsson, T.~Colonius, Instability wave models for the near-field
  fluctuations of turbulent jets, Journal of Fluid Mechanics 689 (2011)
  97--128.
\newblock \href {https://doi.org/https://doi.org/10.1017/jfm.2011.401}
  {\path{doi:https://doi.org/10.1017/jfm.2011.401}}.

\bibitem{sinha2014wavepacket}
A.~Sinha, D.~Rodr{\'\i}guez, G.~A. Br{\`e}s, T.~Colonius, Wavepacket models for
  supersonic jet noise, Journal of Fluid Mechanics 742 (2014) 71--95.
\newblock \href {https://doi.org/https://doi.org/10.1017/jfm.2013.660}
  {\path{doi:https://doi.org/10.1017/jfm.2013.660}}.

\bibitem{towne2015one}
A.~Towne, T.~Colonius, One-way spatial integration of hyperbolic equations,
  Journal of Computational Physics 300 (2015) 844--861.
\newblock \href {https://doi.org/https://doi.org/10.1016/j.jcp.2015.08.015}
  {\path{doi:https://doi.org/10.1016/j.jcp.2015.08.015}}.

\bibitem{rigas2017one}
G.~Rigas, O.~T. Schmidt, T.~Colonius, G.~A. Bres, One way navier-stokes and
  resolvent analysis for modeling coherent structures in a supersonic turbulent
  jet, in: 23rd AIAA/CEAS Aeroacoustics Conference, no. 4046, 2017.
\newblock \href {https://doi.org/https://doi.org/0.2514/6.2017-4046}
  {\path{doi:https://doi.org/0.2514/6.2017-4046}}.

\bibitem{schmidt2016super}
O.~Schmidt, A.~Towne, T.~Colonius, P.~Jordan, V.~Jaunet, A.~V. Cavalieri, G.~A.
  Br{\`e}s, Super-and multi-directive acoustic radiation by linear global modes
  of a turbulent jet, in: 22nd AIAA/CEAS Aeroacoustics Conference, no. 2808,
  2016.
\newblock \href {https://doi.org/https://doi.org/10.2514/6.2016-2808}
  {\path{doi:https://doi.org/10.2514/6.2016-2808}}.

\bibitem{jeun2016input}
J.~Jeun, J.~W. Nichols, M.~R. Jovanovi{\'c}, Input-output analysis of
  high-speed axisymmetric isothermal jet noise, Physics of Fluids 28~(4) (2016)
  047101.
\newblock \href {https://doi.org/https://doi.org/10.1063/1.4946886}
  {\path{doi:https://doi.org/10.1063/1.4946886}}.

\bibitem{towne2017statistical}
A.~Towne, G.~A. Bres, S.~K. Lele, A statistical jet-noise model based on the
  resolvent framework, in: 23rd AIAA/CEAS Aeroacoustics Conference, no. 3706,
  2017.
\newblock \href {https://doi.org/https://doi.org/10.2514/6.2017-3706}
  {\path{doi:https://doi.org/10.2514/6.2017-3706}}.

\bibitem{pickering2019eddy}
E.~M. Pickering, G.~Rigas, D.~Sipp, O.~T. Schmidt, T.~Colonius, Eddy viscosity
  for resolvent-based jet noise models, in: 25th AIAA/CEAS Aeroacoustics
  Conference, no. 2454, 2019.
\newblock \href {https://doi.org/https://doi.org/10.2514/6.2019-2454}
  {\path{doi:https://doi.org/10.2514/6.2019-2454}}.

\bibitem{pickering2020resolvent}
E.~M. Pickering, A.~Towne, P.~Jordan, T.~Colonius, Resolvent-based jet noise
  models: a projection approach, in: AIAA Scitech 2020 Forum, no. 0999, 2020.
\newblock \href {https://doi.org/https://doi.org/10.2514/6.2020-0999}
  {\path{doi:https://doi.org/10.2514/6.2020-0999}}.

\bibitem{kapusta2016numerical}
M.~Kapusta, R.~W. Powers, P.~J. Morris, D.~K. McLaughlin, Numerical simulations
  for supersonic jet noise reduction using fluidic inserts, in: 54th AIAA
  aerospace sciences meeting, no. 0758, 2016.
\newblock \href {https://doi.org/https://doi.org/10.2514/6.2016-0758}
  {\path{doi:https://doi.org/10.2514/6.2016-0758}}.

\bibitem{gaitonde2011coherent}
D.~Gaitonde, M.~Samimy, Coherent structures in plasma-actuator controlled
  supersonic jets: Axisymmetric and mixed azimuthal modes, Physics of Fluids
  23~(9) (2011) 095104.
\newblock \href {https://doi.org/https://doi.org/10.1063/1.3627215}
  {\path{doi:https://doi.org/10.1063/1.3627215}}.

\bibitem{gaitonde2012analysis}
D.~Gaitonde, Analysis of the near field in a plasma-actuator-controlled
  supersonic jet, Journal of Propulsion and Power 28~(2) (2012) 281--292.
\newblock \href {https://doi.org/https://doi.org/10.2514/1.B34289}
  {\path{doi:https://doi.org/10.2514/1.B34289}}.

\bibitem{unnikrishnan2016acoustic}
S.~Unnikrishnan, D.~V. Gaitonde, Acoustic, hydrodynamic and thermal modes in a
  supersonic cold jet, Journal of Fluid Mechanics 800 (2016) 387--432.
\newblock \href {https://doi.org/https://doi.org/10.1017/jfm.2016.410}
  {\path{doi:https://doi.org/10.1017/jfm.2016.410}}.

\bibitem{prasad2019effect}
C.~Prasad, P.~J. Morris, Effect of fluid injection on turbulence and noise
  reduction of a supersonic jet, Philosophical Transactions of the Royal
  Society A 377~(2159) (2019) 20190082.
\newblock \href {https://doi.org/https://doi.org/10.1098/rsta.2019.0082}
  {\path{doi:https://doi.org/10.1098/rsta.2019.0082}}.

\bibitem{prasad2020study}
C.~Prasad, P.~J. Morris, A study of noise reduction mechanisms of jets with
  fluid inserts, Journal of Sound and Vibration (2020) 115331\href
  {https://doi.org/https://doi.org/10.1016/j.jsv.2020.115331}
  {\path{doi:https://doi.org/10.1016/j.jsv.2020.115331}}.

\bibitem{prasad2021steady}
C.~Prasad, P.~J. Morris, Steady active control of noise radiation from highly
  heated supersonic jets, The Journal of the Acoustical Society of America
  149~(2) (2021) 1306--1317.
\newblock \href {https://doi.org/https://doi.org/10.1121/10.0003570}
  {\path{doi:https://doi.org/10.1121/10.0003570}}.

\bibitem{touber2009large}
E.~Touber, N.~D. Sandham, Large-eddy simulation of low-frequency unsteadiness
  in a turbulent shock-induced separation bubble, Theoretical and Computational
  Fluid Dynamics 23~(2) (2009) 79--107.
\newblock \href {https://doi.org/https://doi.org/10.1007/s00162-009-0103-z}
  {\path{doi:https://doi.org/10.1007/s00162-009-0103-z}}.

\bibitem{ranjan2020robust}
R.~Ranjan, S.~Unnikrishnan, D.~Gaitonde, A robust approach for stability
  analysis of complex flows using high-order navier-stokes solvers, Journal of
  Computational Physics 403 (2020) 109076.
\newblock \href {https://doi.org/https://doi.org/10.1016/j.jcp.2019.109076}
  {\path{doi:https://doi.org/10.1016/j.jcp.2019.109076}}.

\bibitem{chakrabarti2021representing}
S.~Chakrabarti, D.~Gaitonde, S.~Unnikrishnan, Representing rectangular jet
  dynamics through azimuthal fourier modes, Physical Review Fluids 6~(7) (2021)
  074605.
\newblock \href
  {https://doi.org/https://doi.org/10.1103/PhysRevFluids.6.074605}
  {\path{doi:https://doi.org/10.1103/PhysRevFluids.6.074605}}.

\bibitem{rodriguez2021near}
D.~Rodriguez, C.~Prasad, D.~Gaitonde, Near-field turbulent structures in
  supersonic rectangular jets using 3d parabolized stability equations, in:
  AIAA Aviation 2021 Forum, no. 2276, 2021.
\newblock \href {https://doi.org/https://doi.org/10.2514/6.2021-2276}
  {\path{doi:https://doi.org/10.2514/6.2021-2276}}.

\bibitem{tinney2008near}
C.~E. Tinney, P.~Jordan, The near pressure field of co-axial subsonic jets,
  Journal of Fluid Mechanics 611 (2008) 175--204.
\newblock \href {https://doi.org/https://doi.org/10.1017/S0022112008001833}
  {\path{doi:https://doi.org/10.1017/S0022112008001833}}.

\bibitem{du2015separation}
Y.~Du, P.~J. Morris, The separation of radiating and non-radiating near-field
  pressure fluctuations in supersonic jets, Journal of Sound and Vibration 355
  (2015) 172--187.
\newblock \href {https://doi.org/https://doi.org/10.1016/j.jsv.2015.06.020}
  {\path{doi:https://doi.org/10.1016/j.jsv.2015.06.020}}.

\bibitem{mancinelli2018hydrodynamic}
M.~Mancinelli, T.~Pagliaroli, R.~Camussi, T.~Castelain, On the hydrodynamic and
  acoustic nature of pressure proper orthogonal decomposition modes in the near
  field of a compressible jet, Journal of Fluid Mechanics 836 (2018) 998--1008.
\newblock \href {https://doi.org/https://doi.org/10.1017/jfm.2017.839}
  {\path{doi:https://doi.org/10.1017/jfm.2017.839}}.

\bibitem{doak1989momentum}
P.~Doak, Momentum potential theory of energy flux carried by momentum
  fluctuations, Journal of sound and vibration 131~(1) (1989) 67--90.
\newblock \href {https://doi.org/https://doi.org/10.1016/0022-460X(89)90824-9}
  {\path{doi:https://doi.org/10.1016/0022-460X(89)90824-9}}.

\bibitem{unnikrishnan2019acoustically}
S.~Unnikrishnan, A.~V. Cavalieri, D.~V. Gaitonde, Acoustically informed
  statistics for wave-packet models, AIAA Journal 57~(6) (2019) 2421--2434.
\newblock \href {https://doi.org/https://doi.org/10.2514/1.J057938}
  {\path{doi:https://doi.org/10.2514/1.J057938}}.

\bibitem{prasad2019modal}
C.~Prasad, P.~J. Morris, Effect of fluid inserts on low order models of jet
  noise reduction, in: 25th AIAA/CEAS Aeroacoustics Conference, no. 2688, 2019.
\newblock \href {https://doi.org/https://doi.org/10.2514/6.2019-2688}
  {\path{doi:https://doi.org/10.2514/6.2019-2688}}.

\bibitem{schmid2010dynamic}
P.~J. Schmid, Dynamic mode decomposition of numerical and experimental data,
  Journal of fluid mechanics 656 (2010) 5--28.
\newblock \href {https://doi.org/https://doi.org/10.1017/S0022112010001217}
  {\path{doi:https://doi.org/10.1017/S0022112010001217}}.

\bibitem{samimy2010acoustic}
M.~Samimy, J.-H. Kim, M.~Kearney-Fischer, A.~Sinha, Acoustic and flow fields of
  an excited high reynolds number axisymmetric supersonic jet, Journal of Fluid
  Mechanics 656 (2010) 507--529.
\newblock \href {https://doi.org/https://doi.org/10.1017/S0022112010001357}
  {\path{doi:https://doi.org/10.1017/S0022112010001357}}.

\bibitem{bogey2010influence}
C.~Bogey, C.~Bailly, Influence of nozzle-exit boundary-layer conditions on the
  flow and acoustic fields of initially laminar jets, Journal of Fluid
  Mechanics 663 (2010) 507--538.
\newblock \href {https://doi.org/https://doi.org/10.1017/S0022112010003605}
  {\path{doi:https://doi.org/10.1017/S0022112010003605}}.

\bibitem{roe1981approximate}
P.~L. Roe, Approximate riemann solvers, parameter vectors, and difference
  schemes, Journal of computational physics 43~(2) (1981) 357--372.
\newblock \href {https://doi.org/https://doi.org/10.1016/0021-9991(81)90128-5}
  {\path{doi:https://doi.org/10.1016/0021-9991(81)90128-5}}.

\bibitem{van1979towards}
B.~Van~Leer, Towards the ultimate conservative difference scheme. v. a
  second-order sequel to godunov's method, Journal of computational Physics
  32~(1) (1979) 101--136.
\newblock \href {https://doi.org/https://doi.org/10.1016/0021-9991(79)90145-1}
  {\path{doi:https://doi.org/10.1016/0021-9991(79)90145-1}}.

\bibitem{pulliam1981diagonal}
T.~H. Pulliam, D.~Chaussee, A diagonal form of an implicit
  approximate-factorization algorithm, Journal of Computational Physics 39~(2)
  (1981) 347--363.
\newblock \href {https://doi.org/https://doi.org/10.1016/0021-9991(81)90156-X}
  {\path{doi:https://doi.org/10.1016/0021-9991(81)90156-X}}.

\bibitem{beam1978implicit}
R.~M. Beam, R.~Warming, An implicit factored scheme for the compressible
  navier-stokes equations, AIAA journal 16~(4) (1978) 393--402.
\newblock \href {https://doi.org/https://doi.org/10.2514/3.60901}
  {\path{doi:https://doi.org/10.2514/3.60901}}.

\bibitem{ffowcs1969sound}
J.~E. Ffowcs~Williams, D.~L. Hawkings, Sound generation by turbulence and
  surfaces in arbitrary motion, Philosophical Transactions of the Royal Society
  of London. Series A, Mathematical and Physical Sciences 264~(1151) (1969)
  321--342.
\newblock \href {https://doi.org/https://doi.org/10.1098/rsta.1969.0031}
  {\path{doi:https://doi.org/10.1098/rsta.1969.0031}}.

\bibitem{samimy2007active}
M.~Samimy, J.-H. Kim, J.~Kastner, I.~Adamovich, Y.~Utkin, Active control of
  high-speed and high-reynolds-number jets using plasma actuators, Journal of
  Fluid Mechanics 578 (2007) 305--330.
\newblock \href {https://doi.org/https://doi.org/10.1017/S0022112007004867}
  {\path{doi:https://doi.org/10.1017/S0022112007004867}}.

\bibitem{kuo2012acoustic}
C.-W. Kuo, J.~Veltin, D.~K. McLaughlin, Acoustic assessment of small-scale
  military-style nozzles with chevrons, Noise Control Engineering Journal
  60~(5) (2012) 559--576.
\newblock \href {https://doi.org/https://doi.org/10.3397/1.3701033}
  {\path{doi:https://doi.org/10.3397/1.3701033}}.

\bibitem{morgan2017further}
J.~Morgan, P.~J. Morris, D.~K. McLaughlin, C.~Prasad, Further development of
  supersonic jet noise reduction using nozzle fluidic inserts, in: 55th AIAA
  Aerospace Sciences Meeting, no. 0683, 2017.
\newblock \href {https://doi.org/https://doi.org/10.2514/6.2017-0683}
  {\path{doi:https://doi.org/10.2514/6.2017-0683}}.

\bibitem{west2015jet}
A.~West, M.~Caraeni, Jet noise prediction using a permeable fw-h solver, in:
  21st AIAA/CEAS Aeroacoustics Conference, no. 2371, 2015.
\newblock \href {https://doi.org/https://doi.org/10.2514/6.2015-2371}
  {\path{doi:https://doi.org/10.2514/6.2015-2371}}.

\bibitem{prasad2019unsteady}
C.~Prasad, P.~J. Morris, Unsteady simulations of fluid inserts for supersonic
  jet noise reduction, in: AIAA Scitech 2019 Forum, no. 0808, 2019.
\newblock \href {https://doi.org/https://doi.org/10.2514/6.2019-0808}
  {\path{doi:https://doi.org/10.2514/6.2019-0808}}.

\bibitem{prasad2020coupled}
C.~Prasad, S.~Hromisin, Coupled les-experimental noise source imaging and
  fluid-thermodynamic mode decomposition of supersonic jets with fluid inserts,
  in: AIAA Scitech 2020 Forum, no. 1000, 2020.
\newblock \href {https://doi.org/https://doi.org/10.2514/6.2020-1000}
  {\path{doi:https://doi.org/10.2514/6.2020-1000}}.

\bibitem{karban_ambiguity_2020}
U.~Karban, B.~Bugeat, E.~Martini, A.~Towne, A.~V.~G. Cavalieri, L.~Lesshafft,
  A.~Agarwal, P.~Jordan, T.~Colonius, Ambiguity in mean-flow-based linear
  analysis, Journal of Fluid Mechanics 900 (2020) R5.
\newblock \href {https://doi.org/https://doi.org/10.1017/jfm.2020.566}
  {\path{doi:https://doi.org/10.1017/jfm.2020.566}}.

\bibitem{bogey2007experimental}
C.~Bogey, S.~Barr{\'e}, V.~Fleury, C.~Bailly, D.~Juv{\'e}, Experimental study
  of the spectral properties of near-field and far-field jet noise,
  International Journal of Aeroacoustics 6~(2) (2007) 73--92.
\newblock \href {https://doi.org/https://doi.org/10.1260/147547207781041868}
  {\path{doi:https://doi.org/10.1260/147547207781041868}}.

\bibitem{van1992bi}
H.~A. Van~der Vorst, Bi-cgstab: A fast and smoothly converging variant of bi-cg
  for the solution of nonsymmetric linear systems, SIAM Journal on Scientific
  and Statistical Computing 13~(2) (1992) 631--644.
\newblock \href {https://doi.org/https://doi.org/10.1137/0913035}
  {\path{doi:https://doi.org/10.1137/0913035}}.

\bibitem{rodriguez2013inlet}
D.~Rodr{\'\i}guez, A.~Sinha, G.~A. Br{\`e}s, T.~Colonius, Inlet conditions for
  wave packet models in turbulent jets based on eigenmode decomposition of
  large eddy simulation data, Physics of Fluids 25~(10) (2013) 105107.
\newblock \href {https://doi.org/https://doi.org/10.1063/1.4824479}
  {\path{doi:https://doi.org/10.1063/1.4824479}}.

\bibitem{schmidt2018spectral}
O.~T. Schmidt, A.~Towne, G.~Rigas, T.~Colonius, G.~A. Br{\`e}s, Spectral
  analysis of jet turbulence, Journal of Fluid Mechanics 855 (2018) 953--982.
\newblock \href {https://doi.org/https://doi.org/10.1017/jfm.2018.675}
  {\path{doi:https://doi.org/10.1017/jfm.2018.675}}.

\bibitem{tissot2017wave}
G.~Tissot, F.~C. Laj{\'u}s~Jr, A.~V. Cavalieri, P.~Jordan, Wave packets and orr
  mechanism in turbulent jets, Physical Review Fluids 2~(9) (2017) 093901.
\newblock \href
  {https://doi.org/https://doi.org/10.1103/PhysRevFluids.2.093901}
  {\path{doi:https://doi.org/10.1103/PhysRevFluids.2.093901}}.

\bibitem{tissot2017sensitivity}
G.~Tissot, M.~Zhang, F.~C. Laj{\'u}s, A.~V. Cavalieri, P.~Jordan, Sensitivity
  of wavepackets in jets to nonlinear effects: the role of the critical layer,
  Journal of Fluid Mechanics 811 (2017) 95--137.
\newblock \href {https://doi.org/https://doi.org/10.1017/jfm.2016.735}
  {\path{doi:https://doi.org/10.1017/jfm.2016.735}}.

\bibitem{Tam2009mach}
C.~K. W.~Tam, Mach wave radiation from high-speed jets, AIAA Journal 47~(10)
  (2009) 2440--2448.
\newblock \href {https://doi.org/https://doi.org/10.2514/1.42644}
  {\path{doi:https://doi.org/10.2514/1.42644}}.

\bibitem{tam1980radiation}
C.~K. Tam, P.~J. Morris, The radiation of sound by the instability waves of a
  compressible plane turbulent shear layer, Journal of Fluid Mechanics 98~(2)
  (1980) 349--381.
\newblock \href {https://doi.org/https://doi.org/10.1017/S0022112080000195}
  {\path{doi:https://doi.org/10.1017/S0022112080000195}}.

\bibitem{tam1984sound}
C.~K. Tam, D.~E. Burton, Sound generated by instability waves of supersonic
  flows. part 1. two-dimensional mixing layers, Journal of Fluid Mechanics 138
  (1984) 249--271.
\newblock \href {https://doi.org/https://doi.org/10.1017/S0022112084000124}
  {\path{doi:https://doi.org/10.1017/S0022112084000124}}.

\bibitem{tam1984sound2}
C.~K. Tam, D.~E. Burton, Sound generated by instability waves of supersonic
  flows. part 2. axisymmetric jets, Journal of Fluid Mechanics 138 (1984)
  273--295.
\newblock \href {https://doi.org/https://doi.org/10.1017/S0022112084000124}
  {\path{doi:https://doi.org/10.1017/S0022112084000124}}.

\bibitem{jovanovic2014sparsity}
M.~R. Jovanovi{\'c}, P.~J. Schmid, J.~W. Nichols, Sparsity-promoting dynamic
  mode decomposition, Physics of Fluids 26~(2) (2014) 024103.
\newblock \href {https://doi.org/https://doi.org/10.1063/1.4863670}
  {\path{doi:https://doi.org/10.1063/1.4863670}}.

\bibitem{cheng2021two}
J.~Cheng, J.~D. Goldschmidt, W.~Shen, L.~Ukeiley, S.~A. Miller, Two-point
  radiation statistics from large-scale turbulent structures within supersonic
  jets, International Journal of Aeroacoustics 20~(3-4) (2021) 254--282.
\newblock \href {https://doi.org/https://doi.org/10.1177/1475472X211005413}
  {\path{doi:https://doi.org/10.1177/1475472X211005413}}.

\bibitem{towne2018spectral}
A.~Towne, O.~T. Schmidt, T.~Colonius, Spectral proper orthogonal decomposition
  and its relationship to dynamic mode decomposition and resolvent analysis,
  Journal of Fluid Mechanics 847 (2018) 821--867.
\newblock \href {https://doi.org/https://doi.org/10.1017/jfm.2018.283}
  {\path{doi:https://doi.org/10.1017/jfm.2018.283}}.

\bibitem{samimy2019reinventing}
M.~Samimy, N.~Webb, A.~Esfahani, Reinventing the wheel: excitation of flow
  instabilities for active flow control using plasma actuators, Journal of
  Physics D: Applied Physics 52~(35) (2019) 354002.
\newblock \href {https://doi.org/https://doi.org/10.1088/1361-6463/ab272d}
  {\path{doi:https://doi.org/10.1088/1361-6463/ab272d}}.

\bibitem{prasad2021extraction}
C.~Prasad, D.~V. Gaitonde, Extraction of fluid thermodynamic modes from the
  mean flow of a supersonic jet, in: AIAA Scitech 2021 Forum, no. 1415, 2021.
\newblock \href {https://doi.org/https://doi.org/10.2514/6.2021-1415}
  {\path{doi:https://doi.org/10.2514/6.2021-1415}}.

\end{thebibliography}

\end{document}